\documentclass[a4paper,usenatbib]{mnras}
\usepackage{aas_macros}
\usepackage{graphicx}
\usepackage{amsmath}
\usepackage{amssymb}
\usepackage[usenames]{xcolor}
\hypersetup{colorlinks=true,citecolor=cyan}

\newcommand{\delc}{\ensuremath{\delta_{\rm c}}}

\newcommand{\avg}[1]{\ensuremath{\left\langle \,#1\, \right\rangle}}

\newcommand{\der}{\ensuremath{{\rm d}}}
\newcommand{\dir}{\ensuremath{\delta_{\rm D}}}
\newcommand{\HT}{\ensuremath{\Theta_{\rm H}}}

\newcommand{\eqn}[1]{equation~\eqref{#1}}
\newcommand{\eqns}[1]{equations~\eqref{#1}}
\newcommand{\fig}[1]{Figure~\ref{#1}}

\newcommand{\ph}[1]{\phantom{#1}}

\newcommand{\be}{\begin{equation}}
\newcommand{\ee}{\end{equation}}
\newcommand{\bea}{\begin{align}}
\newcommand{\eea}{\end{align}}
\newcommand{\Cal}[1]{\ensuremath{\mathcal{#1}}}

\def\Mpc{\, h^{-1} \, {\rm Mpc}}

\date{draft}

\title[Halo formation from halo bias]
      {Constraints on halo formation from cross-correlations with correlated variables} 

\author[E. Castorina, et al.]
{Emanuele Castorina$^{1,2,3}$\thanks{E-mail: ecastorina@berkeley.edu}, Aseem Paranjape$^{4}$ \& Ravi K. Sheth$^{5,6}$\\  
 $^1$ Berkeley Center for Cosmological Physics, University of California, Berkeley, CA 94720, USA\\
$^2$ Lawrence Berkeley National Laboratory, 1 Cyclotron Road, Berkeley, CA 93720, USA\\
 $^3$ SISSA - International School For Advanced Studies,
      Via Bonomea, 265 34136 Trieste, Italy\\
 $^4$     Inter-University Centre for Astronomy and Astrophysics, Ganeshkhind, Post Bag 4, Pune 411007, India\\
 $^5$ Center for Particle Cosmology, University of Pennsylvania, 
      209 S. 33rd St., Philadelphia, PA 19104, USA\\
 $^6$ The Abdus Salam International Center for Theoretical Physics, Strada Costiera, 11, Trieste 34151, Italy
 }

\begin{document}
\pagerange{\pageref{firstpage}--\pageref{lastpage}}

\maketitle 

\label{firstpage}

\begin{abstract}
Cross-correlations between biased tracers and the dark matter field encode information about the physical variables which characterize these tracers. 
However, if the physical variables of interest are correlated with one another, then extracting this information is not as straightforward as one might naively have thought. 
We show how to exploit these correlations so as to estimate scale-independent bias factors of \emph{all} orders 
in a model-independent way. 
We also show that failure to account for this will lead to incorrect conclusions about which variables matter and which do not. 
Morever, accounting for this allows one to use the scale dependence of bias to constrain the physics of halo formation; to date the argument has been phrased the other way around. 
We illustrate by showing that the scale dependence of linear and nonlinear bias, measured on nonlinear scales, can be used to provide consistent estimates of how the critical density for halo formation depends on halo mass. 
Our methods work even when the bias is nonlocal and stochastic, such as when, in addition to the spherically averaged density field and its derivatives, the quadrupolar shear field also matters for halo formation. 
In such models, the nonlocal bias factors are closely related to the more familiar local nonlinear bias factors, which are much easier to measure. 
Our analysis emphasizes the fact that biased tracers are biased because they do not sample fields (density, velocity, shear, etc.) at all positions in space in the same way that the dark matter does. 
\end{abstract}


\section{Introduction}
\label{sec:intro}
\noindent
The point process defined by a galaxy sample is a biased tracer of the underlying dark matter distribution.  This is a consequence of the fact that galaxies populate dark matter haloes, and the haloes themselves are biased tracers \citep[e.g. the review in ][]{cs02}.  Nevertheless, because halo bias is reasonably well understood, the clustering of galaxies in large scale sky surveys encodes information about galaxy formation and also provides competitive constraints on cosmological parameters.  

Halo bias is commonly defined as the ratio of $\avg{\Delta|{\rm halo}}$, the cross correlation of the large scale matter fluctuation $\Delta$ with the halo distribution (at distance $r$), to the cross correlation of $\Delta$ with the matter (rather than halo) distribution (at the same distance $r$):  $\avg{\Delta|{\rm matter}} = \avg{\Delta\delta}$.  Similarly, if $P_{\Delta|h}(k)$ denotes the Fourier transform of $\avg{\Delta|{\rm halo}}$, then the Fourier space bias $b_\delta(k)$ is defined by the ratio $P_{\Delta|h}(k)/P_{\Delta\delta}(k)$.  The main goal of the present note is to show that there is much to be gained if one generalizes this definition by replacing $\Delta$ with any other quantity, say $Y$, associated with the field at distance $r$ from each halo.  The associated Fourier space bias will be $b_y(k)\equiv P_{Y|h}(k)/P_{Y\delta}(k)$.  We will also argue that it is useful to replace the notion of ``bias with respect to the dark matter overdensity field'' with the slightly more general notion of bias with respect to an average over all space whatever the field.  This makes explicit that biased tracers are biased because they do not sample (density, velocity, shear, etc.) fields at all positions in space in the same way that the dark matter does.  Although this was the sense in which bias was initially meant, this particular view of bias has gone out of vogue.  

In Section~2 we show that, in some cases, the $b_y$ may be the same function of $k$ for more than one choice of $Y$.  When this occurs, we show how cross-correlating the halo field with these different choices for $Y$ allows for less model-dependent parameter constraints than were previously possible.  We first show that cross-correlating with different parameters $Y$ provides a simple way of estimating the coefficients which determine the scale dependence of linear bias.  We then show how to combine these to provide information about halo formation.  Finally, we show how to generalize the method to estimate nonlinear bias factors of all orders.  In Section~3 we use Monte-Carlo simulations to show that our approach works well when the density field and its derivatives are the only parameters which matter for halo formation.  Section~4 shows that it continues to work well when the shear field also matters.  A final section summarizes.

\section{Bias of constrained regions}
\subsection{Overdense patches in the initial conditions}
To begin, suppose that we identify those positions in the initial field which, when smoothed on scale $R_p$ have $\delta_p\ge\delta_c$.  This is not a particularly good model of protohaloes -- patches in the initial overdensity fluctuation field which are destined to form haloes -- but it allows us to highlight a number of interesting points before we move on to more realistic models.  

The cross-correlation between the overdensity $\delta_q$ smoothed on scale $R_q\ne R_p$ and $\delta_p$, when it is known that $\delta_p\ge\delta_c$ (the subscript $c$ denotes `critical'; it does not denote a scale $R_c$) is given by 
\begin{equation}
 \avg{\!\delta_q|{\cal C}_p\!} = 
 \int_{\delta_c}^\infty {\rm d}\delta_p\,p(\delta_p)
          \,\frac{\langle\delta_q|\delta_p\rangle}{p({\cal C}_p)},
\end{equation}
where ${\cal C}_p \equiv (\delta_p\ge\delta_c)$.  
Since 
\begin{equation}
 \avg{\!\delta_q|\delta_p\!} \equiv
 \frac{\avg{\!\delta_q\delta_p\!}}{\avg{\!\delta_p\delta_p\!}}\,\delta_p,
\end{equation}
we have that 
\begin{equation}
 \avg{\!\delta_q|{\cal C}_p\!} = b_{\rm L}(\nu)\,\avg{\!\delta_q\delta_p\!},
 \label{dthreshold}
\end{equation}
where 
\be
 b_{\rm L}(\nu) \equiv \frac{\nu}{\delta_c}
             \,\frac{{\rm e}^{-\nu^2/2}/\sqrt{2\pi}}{{\rm erfc}(\nu/\sqrt{2})/2}
 \quad{\rm with}\quad
 \nu\equiv \frac{\delta_c}{\sqrt{\avg{\!\delta_p\delta_p\!}}}.
 \label{bpatch}
\ee
Evidently, the cross-correlation between the positions identified as satisfying the constraint ${\cal C}_p$ on scale $R_p$ and the field smoothed on scale $R_q$ is linearly proportional to the cross-correlation of the field itself when smoothed on the two scales $R_p$ and $R_q$.  \cite[][show that this is a generic feature of local Lagrangian bias models such as this one.]{fs12}  The constant of proportionality defines the linear bias factor.  It is a complicated function of the threshold, but is otherwise just a number which is independent of the scale $R_q$.  

For what follows, it is useful to also study the Fourier transform of the cross-correlation, which is given by 
\be
 P_{\delta|{\cal C}}(k) = b_{\rm L}(\nu)\,P_{\delta\delta}(k)\, W(kR_q)\, W(kR_p),
\ee
where the $W$ denote the Fourier transforms of the smoothing windows.  Although the most natural choice has $W(x)$ being the same function of $x$ on both scales $R_p$ and $R_q$, the formalism is sufficiently general that different functional forms $W$ for the different scales are permitted:  we could have written the two $W$s in the expression above as $W_q(kR_q)$ and $W_p(kr_p)$.

In the example above, $\delta_p$ is special, since the constraint ${\cal C}_p$ depends only on $\delta$ on the scale $R_p$.  This raises the question of whether or not we can infer, from cross-correlation measurements alone, that the constraint depends only on the physical parameter $\delta$ smoothed on scale $R_p$.  (In anticipation of the discussion in the next section, it may be useful to think of $y_p\equiv \der\delta_p/\der\!\ln\avg{\!\delta_p\delta_p\!}$ as a specific example of another variable on scale $R_p$.)  For example, one might measure the cross-correlation $\avg{\!\delta_q\delta_r\!}$, where $\delta_r$ denotes the field smoothed on scale $R_r$ and the average is over all positions, not just those which have $\delta_p\ge \delta_c$.  Then the ratio 
\be
 \frac{\avg{\!\delta_q|{\cal C}_p\!}}{\avg{\!\delta_q\delta_r\!}}
 = b_{\rm L}(\nu) \,\frac{\avg{\!\delta_q\delta_p}}{\avg{\!\delta_q\delta_r\!}}.
\ee
If one repeated this analysis for a range of $R_r$ and $R_q$, one would find that the real-space ratio was a function of $R_q$ except when $R_r = R_p$, when the ratio would equal the constant $b_{\rm L}(\nu)$.  One might then conclude that $R_p$ was special.  Analysis of the ratio of the associated Fourier space quantities would lead to a similar conclusion, since 
\be
 \frac{P_{\delta|{\cal C}}(k)}{P_{\delta\delta}(k) W(kR_q)W(kR_r)}
 = b_{\rm L}(\nu)\, \frac{W(kR_p)}{W(kR_r)}
\ee
is a function of $k$ except when $R_r=R_p$, when it equals $b_{\rm L}(\nu)$.  

This is not quite as trivial as it sounds of course, since one is also free to vary the shape of the smoothing filter -- which may not be known {\em a priori} --  but the logic of the approach is clear.  In this case, the Fourier space approach is more illuminating, since if $W(kR_r) \to 1$ when $R_r\to 0$ (as is true for the uncompensated filters which are commonly used in this context), then the Fourier transform of $\avg{\!\delta_q\delta_r\!}\to P_{\delta\delta}(k)\, W(kR_q)$.  In this limit, the ratio $P_{\delta|{\cal C}}(k)/[P_{\delta\delta}(k)\, W(kR_q)] = b_{\rm L}(\nu)\, W(kR_p)$, and since $b_{\rm L}$ does not depend on $k$, one has determined the shape of $W(kR_p)$.  

Of course, in practice one may not know that $\delta$ is the relevant physical variable, so one may not know that the relevant cross-correlation to study is $\avg{\!\delta_q\delta_r\!}$ rather than, say, $\avg{\!\delta_q\,y_r\!}$.  However, scale independence is a useful guide here too.  For example, if $W(kR_r) = \exp(-k^2R_r^2/2)$ and $y_r = \der \delta_r/\der\ln\avg{\!\delta_r^2\!}$
then $\avg{\!\delta_q\, y_r}$ will differ from $\avg{\!\delta_q\delta_r\!}$ by a term that is proportional to $k^2$.  So, even if $R_r=R_p$, the ratio $P_{\delta|{\cal C}}(k)/[P_{y\delta}(k)W(kR_q)]$ will be $k$-dependent.  

One might have thought that since ${\cal C}_p$ depends only on $\delta_p$, cross-correlating with the overdensity on another scale is a natural choice, and this is why the analysis yields a `linear', `scale-independent' bias factor.  This would be a concern, since it suggests that to measure the linear bias factor, one needs to know {\em a priori} which variable (or variables) were important for determining what the constraint was.  

To see why this is incorrect, suppose that, instead of cross-correlating the special positions with $\delta_q$, one cross-correlated with some other variable $y_q$ defined on scale $R_q$.  (Again, it may be useful to think of $y_q\equiv \der\delta_q/\der\!\ln\avg{\!\delta_q\delta_q\!}$ as a specific example.)  Then the same logic as above yields 
\be
 \avg{\!y_q|{\cal C}_p\!} = b_{\rm L}(\nu)\,\avg{\!y_q\delta_p\!},
\ee
with associated Fourier transform 
\be
 P_{y|{\cal C}}(k) = b_{\rm L}(\nu)\,P_{y\delta}(k)\, W(kR_q)\, W(kR_p).
\ee
In this case, because one has cross-correlated with $y$, one would study the ratio 
\be
 \frac{\avg{\!y_q|{\cal C}_p\!}}{\avg{\!y_q\delta_r\!}}
 = b_{\rm L}(\nu)\,\frac{\avg{\!\delta_q\delta_p}}{\avg{\!y_q\delta_r\!}}
\ee
or 
\be
 \frac{P_{y|{\cal C}}(k)}{P_{y\delta}(k)W(kR_q)W(kR_r)} = b_{\rm L}(\nu)\, \frac{W(kR_p)}{W(kR_r)}
\ee
for a range of $R_r$, and would find the ratio to be independent of $R_q$ or $k$ only if $R_r = R_p$.  Notice that the bias factor is the {\em same} for all choices of $y_q$.  In particular, $b_{\rm L}$ is the same as when $y_q=\delta_q$, showing that there is nothing special about $\delta_q$.  I.e., cross-correlating with {\em any} variable which correlates with $\delta_p$ will yield the same `linear' bias factor -- this generalizes the argument in \cite{fs12}. 
The lesson is that the bias factor itself contains information about the important variable (i.e., that it is $\delta_p$ which must exceed a threshold), and this information can be obtained from {\em any} quantity which correlates with it.  

Since it is not necessary that one choose the same variable smoothed on another scale, one no longer need know {\em a priori} that it was $\delta_p$ which mattered.  This is a significant simplification.  There are, of course, signal-to-noise issues, since one estimates the bias by dividing $\avg{\!y_q|\delta_p\ge\delta_c\!}$ by $\avg{\!y_q\delta_p\!}$, so it is important that $\avg{\!y_q\delta_p\!}$ be far from zero.  This suggests one should choose $y_q$ for which $\avg{\!y_q\delta_p\!}$ is larger, so in this sense knowing that $\delta_p$ matters is important.  But note that this is a measurement issue, and not one of principle.  In principle, one can determine the linear bias factor $b_{\rm L}$ having much less knowledge of which variable actually matters (in this case $\delta_p$) than one might naively have thought.  

Finally, from the analysis above, it should be clear that there was nothing special about $\delta$.  For example, one might instead have set ${\cal C}_p = (y_p\ge y_c)$ for some other physical parameter $y$.  Then, the analysis above implies that it is cross-correlations with $y$ which will play a fundamental role; cross-correlations with $\delta$ will be $k$-dependent; and the linear scale independent bias $b_{\rm L}$ factor will be a function of $\nu = y_c/\avg{\!y_p^2\!}^{1/2}$.  This is the sense in which it is the bias with respect to unconstrained averages (of the $y$ field, in this case) which is simple and linear; the bias with respect to the $\delta$ field will appear more complicated.  

\subsection{When more than one variable matters}
The previous subsection studied constrained averages in which the only constraint was on the overdensity field on one scale (or, more generally, on one variable on one scale).  The excursion set approach asserts that protohalo patches satisfy more complex constraints.  In particular, in the upcrossing approximation associated with the simplest excursion set approach, protohaloes of mass $m$ are associated with regions of size $R_p\propto m^{1/3}$ in the initial conditions where $\delta_p\ge \delta_c(R_p)$ and $\delta(R_p+\Delta R)<\delta_c(R_p+\Delta R)$ \citep{bcek91, ms12}.  
Notice that this is like requiring that both $\delta$ and its derivative exceed a threshold; it is in this sense that this is a more complex model than the previous one.  

In this approximation, the cross-correlation between protohaloes and some other quantity $Y$ in the initial fluctuation field is
\be
 \avg{Y|{\cal C}_p}
   = f(s)^{-1}\int_{\delta_c'}^\infty \der v\,(v-\delta_c')\,p(\delta_c,v)\,\avg{Y|\delta_c,v}
 \label{eq:avYup}
\ee
where $v\equiv \der\delta_p/\der s$ with $s\equiv\avg{\delta^2_p}$, and 
\be
 f(s) = \int_{\delta_c'}^\infty \der v\,(v-\delta_c')\,p(\delta_c,v),
\label{eq:fup}
\ee
where $\delta_c'\equiv \der\delta_c(s)/\der s$.  

Setting $x\equiv v/\sqrt{\avg{v^2}} \equiv v/\sqrt{s_v}$ and $\gamma\equiv \avg{v\delta_p}/\sqrt{\avg{\delta_p^2}\avg{v^2}} = (2\sqrt{ss_v})^{-1}$ yields 
\be
 \avg{Y|\delta_c,v} =
  \avg{Y\delta_p}\frac{\nu - \gamma x}{\sqrt{s}\,(1-\gamma^2)}
  + \avg{Yv} \frac{x - \gamma\nu}{\sqrt{s_v}\,(1-\gamma^2)}.
\ee
The terms $\avg{Y\delta_p}$ and $\avg{Yv}$ depend differently on scale because 
$\avg{Yv} = \der\avg{Y\delta_p}/\der s$.  Since we define bias as the ratio of the halo-$Y$ correlation with respect to $\avg{Y\delta_p}$, the term which is proportional to $\avg{Y\delta_p}$ will be independent of scale.  To match notation with previous work, we call this scale-independent term $b_{10}$.  Note that its amplitude is set by the variables which determine halo formation (in this case $\delta_p$ and $v$) and {\em not} by the variable $Y$ with which we chose to cross correlate the halo field.  In the example above, 
\be
 b_{10}  = \frac{\delta_c - \avg{v|\Cal{C}_p}/2s_v}{s\,(1-\gamma^2)}
\ee
where $\avg{v|\Cal{C}_p} = f(s)^{-1}\int_{\delta_c'}^\infty\der v\,v(v-\delta_c')p(\delta_c,v)$.
A little algebra then shows that  
\be
 b_1 \equiv \frac{\avg{Y|{\cal C}_p}}{\avg{Y\delta_p}} = b_{10}
           + \epsilon_{Y\delta_p}\, b_{11},
 \label{byr}
\ee
where 
\be
 \epsilon_{Y\delta_p}\equiv 2\,\frac{\der\ln\avg{Y\delta_p}}{\der\ln s}
 \label{epsilon}
\ee
and 
\be
 b_{11} \equiv \frac{\delta_c}{s} - b_{10}.
 \label{b11fromb10}
\ee
Whereas $b_{11}$, like $b_{10}$, depends only on the quantities which determine ${\cal C}_p$ and not on the $Y$ field, the term it multiplies is scale dependent in real space and $k$-dependent in Fourier space.  This is particularly easy to see if we write the ratio of the constrained to unconstrained power spectra:
\be
 \frac{P_{Y|{\cal C}}(k)}{P_{Y\delta}(k)} = b_{10}\,W(kR_p) 
  + 2 \frac{\der W(kR_p)}{\der\ln s}\,b_{11},
 \label{byk}
\ee
where we have assumed that the $\delta$ in $P_{Y\delta}(k)$ was unsmoothed (i.e. it is $\delta_r$ when $R_r\to 0$), since neither the shape of $W(kR_p)$ nor the scale $R_p$ are known {\em a priori}.  E.g., if $W(kR_p) = \exp(-k^2R_p^2/2)$, then the right hand side equals $b_{10}$ plus a term which is proportional to $R_p^2k^2$ all multiplied by $W(kR_p)$.

Notice that the right hand side of equation~(\ref{byk}) is independent of anything to do with $Y$.  (Things are not as simple in real-space, since correlations with $Y$ appear on the right-hand side of equation~\ref{byr}, a fact we exploit shortly.)  This shows explicitly that, even though the constraints ${\cal C}_p$ are more complicated than in the previous section, and they result in $k$-dependent bias (previously the ratio of constrained to unconstrained correlations was independent of $k$), one gets the {\em same} $k$-dependent bias by cross-correlating the protohaloes with {\em any} field $Y$ which is correlated with one of the variables which determine the constraints ${\cal C}_p$.  

Although this is a straightforward generalization of what we found in the previous section -- so the same caveats about signal-to-noise apply here too -- it has an important consequence.  This is because the traditional estimator of $b_{10}$ sets $Y = \delta_q$ with $R_q\gg R_p$, and measures the ratio with respect to \avg{\!Y\delta_p\!}.  The choice of large $R_q$ is motivated by the fact that the $b_{10}$ term dominates in this limit, as can be seen by studying the $kR_p\ll 1$ limit of the Fourier space expressions.  But the decision to set $Y=\delta$ rather than some other field is motivated by the fact that $\Cal{C}_p$ depends on $\delta_p$, and the $b_{10}$ term is the prefactor of the $\avg{\!\delta_q\delta_p\!}$ correlation.  Since $b_{11}$ is the prefactor of the term involving the correlations with $\der\delta_p/\der s$ (the other variable which matters for ${\cal C}_p$), one might naively have thought that setting $Y=\der\delta_q/\der\avg{\!\delta_q^2\!}$ and looking for the $k$-independent part is the way to isolate $b_{11}$.  The analysis above shows that if one defines the bias by ratioing to $\avg{\!Y\delta_p\!}$ as is conventional, then this will not work:  the $k$-independent part will still be $b_{10}$.  

To get $b_{11}$ from scale-independence, equation~(\ref{byr}) shows that one must ratio to $\avg{\!Y\,\der\delta_p/\der\ln s\!}$ instead.  Doing so will make the term which multiplies $b_{10}$ increase at $k\ll R_p$, so it is only at larger $k$ that the $b_{11}$ term may dominate.  However, at these larger $k$, the term which multiplies $b_{11}$ in equation~(\ref{byk}) will show some scale dependence because we are potentially in the regime where the $k$-dependence of $W(kR_p)$ matters (it may help to think of a Gaussian filter here).  As a result, $k$-independent bias will not be as trivial to identify as for $b_{10}$.  Nevertheless, the distinction between these two procedures -- which we may write schematically as $\avg{\!v_q|{\cal C}_p\!}/\avg{\!v_q\delta_p\!}$ and $\avg{\!\delta_q|{\cal C}_p\!}/\avg{\!\delta_qv_p\!}$ -- has led to some confusion in the literature, as we discuss in the next section.  

We turn, therefore, to the question of estimating $b_{11}$ in some other way.  We have already noted that the traditional estimator of $b_{10}$ sets $Y = \delta_q$ with $R_q\gg R_p$.  
\cite{mps12} note that $b_{11}$ is then simply determined by equation~(\ref{b11fromb10}).
However, equation~(\ref{b11fromb10}) assumes prior knowledge of the physics of collapse -- in this case, that $\delta_c$ is the same constant for all values of $s$, and that $\delta$ is the important variable.  Hence, it is interesting to see if we can determine $b_{11}$ without prior knowledge of how $\delta_c$ depends on $s$.  

One possibility is to fit the Fourier space ratio using equation~(\ref{byk}).  This requires prior knowledge that only $W(x)$ and $\der W(x)/\der x$ matter, which boils down to knowing that $\delta$ and its derivative both mattered for ${\cal C}_p$.  Since $W$ is usually close to a tophat with slightly rounded edges, it is relatively straightforward to fit for the scale $R_p$ as well as the bias factors $b_{10}$ and $b_{11}$ \citep[e.g.][]{Chan16}.  

In real space, equation~(\ref{byr}) shows that we can isolate $b_{11}$, the term which controls the scale dependence, if we compute the (real space) bias for two choices of $Y$, and subtract the two expressions.  (This cannot be done in Fourier space, since the right hand side of equation~\ref{byk} is independent of $Y$, so it would yield zero!)  Doing so yields
\be
 \frac{\avg{Y_1|{\cal C}_p}}{\avg{Y_1\delta_p}} - 
 \frac{\avg{Y_2|{\cal C}_p}}{\avg{Y_2\delta_p}} = 2\,b_{11}
  \frac{\der\ln[\avg{Y_1\delta_p}/\avg{Y_2\delta_p}]}{\der\ln s},
 \label{diffb1s}
\ee
which can be rearranged to provide a practical estimator for $b_{11}$ which \emph{does not require} prior knowledge of $\delta_c(s)$.  With $b_{11}$ in hand, one can go on to estimate $b_{10}$ from equation~(\ref{byr}), again without prior knowledge of $\delta_c(s)$.  In fact, equation~(\ref{b11fromb10}) shows that if one adds the estimators of $b_{10}$ and $b_{11}$, then this furnishes an estimate of $\delta_c(s)$.  

Note that the $Y_i$ can be $\delta(R_i)$ on two different scales $R_i$, or one can be $\delta_i$ and the other $\der\delta_j/\der\avg{\!\delta_j^2\!}$, and so on.  In practice, some of these combinations will provide higher signal-to-noise estimators of $b_{11}$ than others.  For example, if $Y_1$ is the first variable, then the natural choice for the second variable $Y_2$ would actually be $Y_2 - \langle Y_2|Y_1\rangle$.  But the point is that each pair furnishes an estimate of $b_{11}$ (and $b_{10}$) which does not depend on any assumptions about $\delta_c(s)$.  However, they all require prior knowledge of $W(kR_p)$ (to estimate the $ \avg{\!Y_i\delta_p\!}$ terms).

Of course, we could have played this the other way round.  Had we started from $\avg{\!Y|{\cal C}_p\!}/\avg{\! Yv\!}$ rather than $\avg{\!Y|{\cal C}_p\!}/\avg{\! Y\delta_p\!}$, then the constrained and unconstrained averages for the two different $Y_i$ could be rearranged to provide an estimator of $b_{10}$ rather than $b_{11}$.  Equation~(\ref{byr}) then yields $b_{11}$, and equation~(\ref{b11fromb10}) yields $\delta_c(s)$. 

For what follows, it is useful to formulate the analysis above in matrix notation.  We have argued that, for any two variables, $Y_1$ and $Y_2$, 
\be
 \begin{pmatrix}
  b_{2Y_1}\\
  b_{2Y_2}
 \end{pmatrix}
 = 
 \begin{pmatrix}
  1 & \epsilon_{Y_1\delta_p} \\
  1 & \epsilon_{Y_2\delta_p} 
 \end{pmatrix}
 \begin{pmatrix}
  b_{10}\\
  b_{11}
 \end{pmatrix} ,
\ee
Therefore, estimates of the left hand side, say $\widehat{b}_{1Y_1}$ and $\widehat{b}_{1Y_2}$, can be turned into estimates of $b_{10}$ and $b_{11}$ because 
\be
 \begin{pmatrix}
  b_{10}\\
  b_{11}
 \end{pmatrix}
 = 
 \begin{pmatrix}
  1 & \epsilon_{Y_1\delta_p} \\
  1 & \epsilon_{Y_2\delta_p} 
 \end{pmatrix}^{-1}
 \begin{pmatrix}
  \widehat{b}_{1Y_1}\\
  \widehat{b}_{1Y_2}
 \end{pmatrix} .
 \label{2x2}
\ee

\subsection{Generalization to higher order bias}
Define $y\equiv Y/\langle Y^2\rangle^{1/2}$.  Then equation~(\ref{eq:avYup}) can be generalized to 
\begin{align}
 b_n &\equiv (-1)^n\frac{\langle Y^2\rangle^{n/2}}{\langle Y\delta_p\rangle^n}
 \avg{H_n(y)|{\cal C}_p}\notag\\
 &= \int_{\delta_c'}^\infty \der v\,(v-\delta_c')\,
     \left(\frac{\partial}{\partial\delta_c} + \frac{\langle Yv\rangle}{\langle Y\delta_p\rangle}\frac{\partial}{\partial v}\right)^n\frac{p(\delta_c,v)}{f(s)},\notag\\
 &= \sum_{j=0}^n {n\choose j}\, b_{nj} \,\epsilon_{Y\delta_p}^j\,,
 \label{eq:bny}
\end{align}
where the $H_n(y)$ are the probabilist's Hermite polynomials.  The second line follows from recognizing that all the steps in Appendix~A2 of \cite{mps12} go through unchanged if one replaces their $\delta_0$ (our $\Delta$) with any other variable $Y$.  Of course, one must recognize that their $S_\times\to\langle Y\delta_p\rangle$ and their $\epsilon_\times\to\epsilon_{Y\delta_p}$.  

The key point here is that the derivatives with respect to $\delta_c$ and $v$ generate the {\em same} bias coefficients $b_{nj}$ for {\em all} choices of $Y$.  E.g., for $n=2$, 
\be
\label{eq:b2}
 b_2 = b_{20} + 2\,b_{21}\,\epsilon_{Y\delta_p} + b_{22}\,\epsilon_{Y\delta_p}^2.
\ee
We can estimate these from the data, with no knowledge of $\delta_c$, using the same approach as for $b_1$, except that now we need three rather than two different $Y$s.  Following the logic of the previous section, we first write 
\be
 \begin{pmatrix}
  b_{2Y_1}\\
  b_{2Y_2}\\
  b_{2Y_3}
 \end{pmatrix}
 = 
 \begin{pmatrix}
  1 & 2\, \epsilon_{Y_1\delta_p} & \epsilon^2_{Y_1\delta_p} \\
  1 & 2\, \epsilon_{Y_2\delta_p} & \epsilon^2_{Y_2\delta_p} \\
  1 & 2\, \epsilon_{Y_3\delta_p} & \epsilon^2_{Y_3\delta_p} 
 \end{pmatrix}
 \begin{pmatrix}
  b_{20}\\
  b_{21}\\
  b_{22}
 \end{pmatrix} ,
\ee
which shows that 
\be
 \begin{pmatrix}
  b_{20}\\
  b_{21}\\
  b_{22}
 \end{pmatrix}
 = 
 \begin{pmatrix}
  1 & 2\, \epsilon_{Y_1\delta_p} & \epsilon^2_{Y_1\delta_p} \\
  1 & 2\, \epsilon_{Y_2\delta_p} & \epsilon^2_{Y_2\delta_p} \\
  1 & 2\, \epsilon_{Y_3\delta_p} & \epsilon^2_{Y_3\delta_p} 
 \end{pmatrix}^{-1}
 \begin{pmatrix}
  \widehat{b}_{2Y_1}\\
  \widehat{b}_{2Y_2}\\
  \widehat{b}_{2Y_3}
 \end{pmatrix} .
 \label{3x3}
\ee
We describe our estimators for the various $\widehat{b}_{2Y}$ in the next section.  The consistency relation -- the analogue of equation~(\ref{b11fromb10}) -- becomes 
\be
 b_{20} + 2b_{21} + b_{22} = H_2\left(\delta_c/\sqrt{s_p}\right)/s_p.
 \label{b2todc}
\ee  
Written this way, the generalization to $b_n$ is trivial:  One makes $n+1$ estimates of $b_{nY_i}$ and then inverts the associated matrix (whose $ij$th element is $\epsilon^j_{i\times}\, n!/j!(n-j)!$).  The consistency relation is 
\be
 \sum_{j=0}^n {n\choose j} b_{nj} = H_n\left(\delta_c/\sqrt{s_p}\right)/s_p^{n/2}.
 \label{bn->dc}
\ee  
Finally, although we do not exploit this in what follows, it is worth noting that, with these replacements, equation~(34) of \cite{mps12} expresses the conditional distribution $f(s|y)$ as a Taylor series in $y$ (these replacements must also be made in their expression for the quantity they call $\bar{\bmath{c}}$).  Previous work has only really considered $f(s|\Delta)$ as a Taylor series in $\Delta$.  It is in this sense that our analysis generalizes the usual notion of bias from being with respect to the large scale environment to any other variable.

\begin{figure}
 \centering
 \includegraphics[width=0.9\hsize]{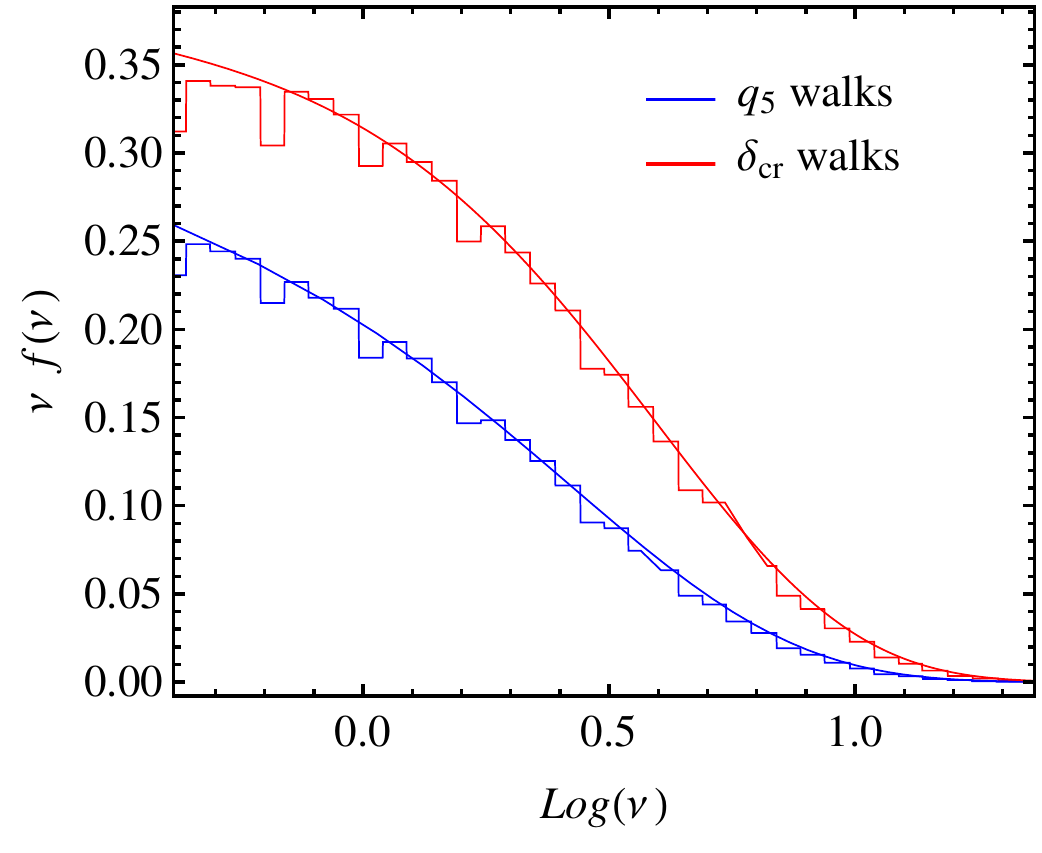}
 \caption{First crossing distribution of a barrier of height $\delta_c\,(1 + q/q_c)$ for Gaussian smoothing of walks having a $\Lambda$CDM $P(k)$.  Histograms show Monte-Carlo first crossing distributions for $q_c=8$ (lower) and $\infty$ (upper); curves show the analytic prediction.  The agreement indicates that our upcrossing approach is a good approximation, even for the stochastic barrier model which we consider in the next section.}
 \label{fig:fs}
\end{figure}

\section{Numerical tests}
\label{subsec:walks}
We now use measurements in Monte-Carlo realizations of random walks crossing a barrier to illustrate some of these ideas.  In all cases, walks were generated using a $\Lambda$CDM $P(k)$ (Planck cosmology) with correlations between steps being due to Top-Hat smoothing filters.  The upper histogram in Figure~\ref{fig:fs} shows the first crossing distribution of a barrier of height $\delta_c$ (the quantity $\nu\equiv\delta_c/\sqrt{s}$ where $s$ is the first crossing scale $s$).  The curve passing through shows the prediction which is based on the assumption that it is only the walk height and its derivative on the upcrossing scale which matter.  This agreement is important, since the bias formulae which we wish to test also result from this `upcrossing' approximation, our \eqn{eq:fup}.  

The lower histogram and associated curve show results when more than one independent variable determines if the barrier $\delta_c$ has been crossed (a model we describe in more detail in Section \ref{subsec:qwalks}).

\begin{figure}
 \centering
 \includegraphics[width=0.9\hsize]{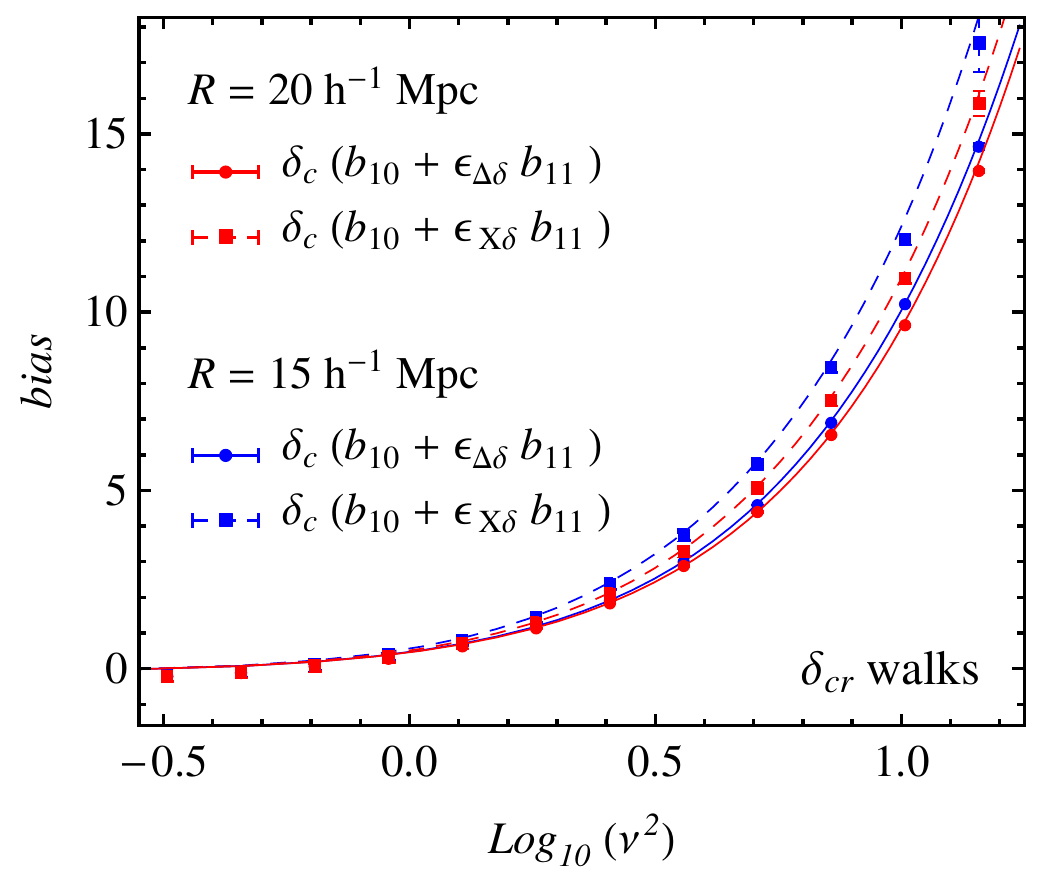}
 \caption{Bias factor $b_1$, shown as a function of $\nu^2\equiv\delta_c^2/s$, estimated from the mean value of $Y$ on the scale $R$ when the first crossing scale was $s>S(R)$ (symbols). Filled circles show measurements for  $Y_1=\Delta$ for two different smoothing scales $R$, while filled squares are for $Y_2=\,d\Delta/d\ S  = X$.  Curves show equation~(\ref{byr}), where $\epsilon_{Y_i\delta}\equiv 2\der\ln\avg{Y_i\Delta}/\der\ln S$. }
 \label{fig:b1s}
\end{figure}

With a slight abuse of notation, let $b_Y$ denote the mean value of $Y$ when smoothed on scale $S$ around a walk which first crossed the barrier on scale $s\ne S$ (typically $s>S$), in units of the unconstrained cross correlation between $Y$ and $\delta$.  I.e., $b_Y$ is $b_1$ when the cross-correlation is with the variable $Y$.  Then, we can estimate it from the walks by measuring \citep{szalay88,mps12}
\be
 \widehat{b}_Y = \frac{1}{N}\sum_{\alpha=1}^N \frac{Y_\alpha}{\langle Y\delta_p\rangle}
 \label{eq:b1_hat}
\ee
where the sum is over those walks which first crossed $\delta_c(s)$ on scale $s$.  Note that $\widehat{b}_Y$ will be a function of the scale on which $Y$ was estimated, as well as of the constraint:  $s$ and $\delta_c(s)$.  The denominator shows that it requires knowledge of the fact that the scale $s$ is special.  In contrast, $\widehat{b}_Y\,\langle Y\delta_p\rangle/\langle Y^2\rangle$, the quantity studied by \cite{mps12}, does not require this knowledge.

Figure~\ref{fig:b1s} shows $\widehat{b}_Y$ for two choices of $Y$:  $Y_1=\Delta$ (filled circles) and $Y_2=\der\Delta/\der\ln S$ (filled squares), and two choices of smoothing scale $R$ (as indicated).  
In all cases, $\delta$ is in units of $\delta_c$ so the normalization factor in equation~(\ref{eq:b1_hat}) is $\avg{Y\delta_p}/\delta_c$, and we show $s$ as $\nu^2\equiv \delta_c^2/s$.  

First, notice that both choices of $Y$ result in similar signals, and this signal is rather similar for different $R$.  To show that we understand the small differences between $Y$'s, as well as the (smaller!) differences between smoothing scales, the  curves show equation~(\ref{byr}) for the appropriate values of $Y$, $R$ and $s$.  Since we have normalized our measurements by $\avg{Y\delta_p}/\delta_c$ rather than $\avg{Y\delta_p}$ itself, the curves actually show the right hand side of equation~(\ref{byr}) multiplied by $\delta_c$.

\begin{figure}
 \centering
 \includegraphics[width=0.9\hsize]{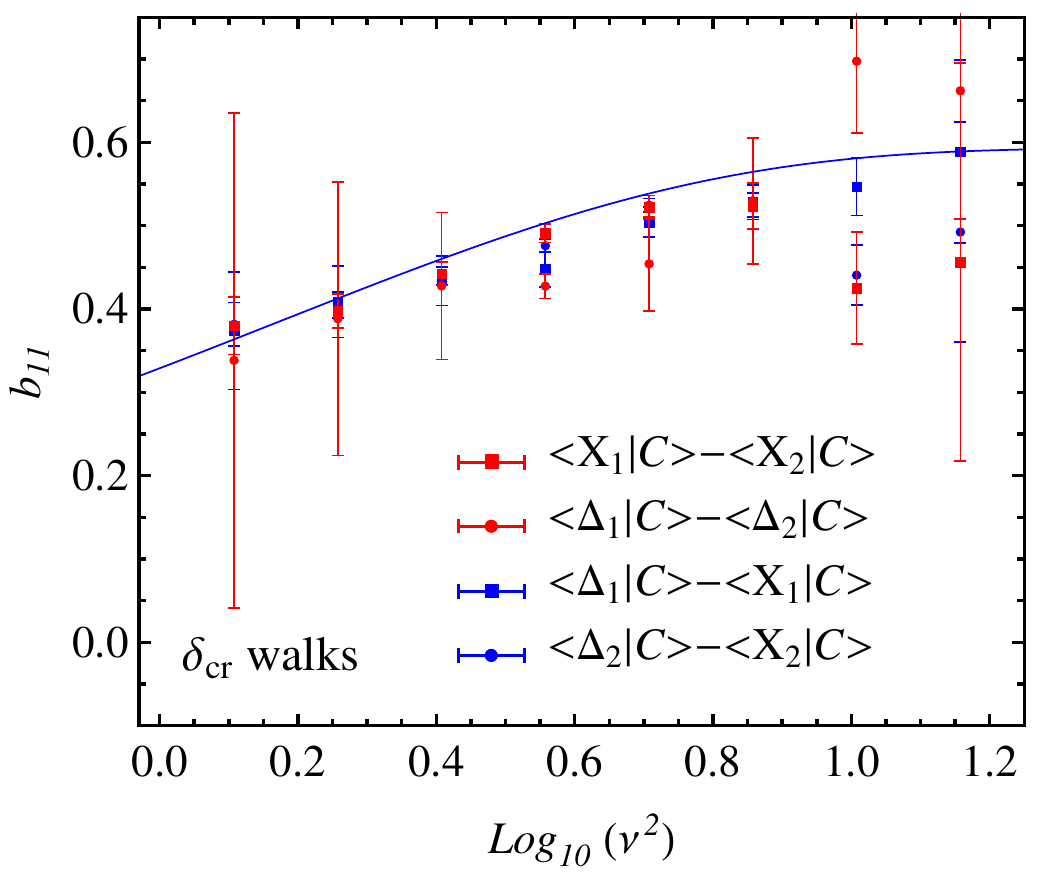}
 \caption{Bias factor $b_{11}$ estimated using different combinations of $Y_1=\Delta$ and $Y_2=\,{\rm d}\Delta/{\rm d}S$, on the scales $R=15 \Mpc$ and/or $R=20\Mpc$.  Curve shows the expected value (equation~\ref{diffb1s}).  }
 \label{fig:b11}
\end{figure}

In all cases the curves pass through the associated symbols, indicating that both choices of $Y$ estimate $b_{10}$ plus a small correction factor which depends on $b_{11}$ (and slightly on scale $R$).  Whereas it is well-known that the cross-correlation with $\Delta$ yields what is essentially an estimate of $b_{10}$, this shows explicitly that the cross-correlation with $2\,\der\Delta/\der\ln S$ also estimates $b_{10}$ (rather than $b_{11}$).  This illustrates one of our main results.

Previous work \cite[e.g.][]{mps12} turns this estimate of $b_1$ into an estimate of $b_{10}$ using the consistency relation (equation~\ref{b11fromb10}).  However, we argued that one can estimate $b_{11}$ directly -- with no assumptions about $\delta_c$ -- by differencing $b_1$ estimates derived from two different choices of $Y$, or the same $Y$ at different smoothing scales.  E.g., if $Y_1=\Delta$
and $Y_2 = X \equiv {\rm d}\Delta/{\rm d}S$, then
\begin{equation}
 \widehat{b}_{11}(s) = \frac{1}{N}\sum_{\alpha=1}^N
 \frac{\Delta_\alpha/\langle\Delta\delta\rangle - X_{\alpha}/\langle X\delta\rangle}{\epsilon_{\Delta\delta}-\epsilon_{X\delta}}
 \label{eq:b11_hat}
\end{equation}
where $\epsilon_{Y\delta}$ was defined in equation~(\ref{epsilon}).

\begin{figure}
 \centering
 \includegraphics[width=0.9\hsize]{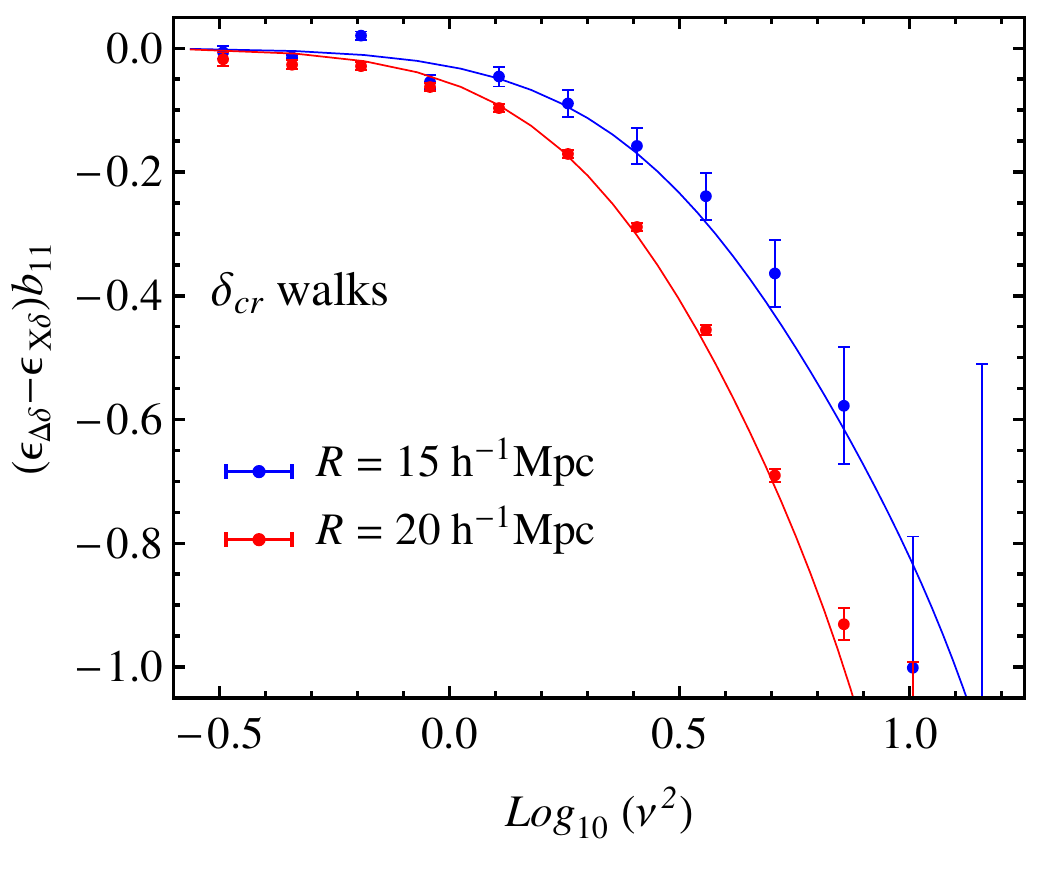}
 \includegraphics[width=0.9\hsize]{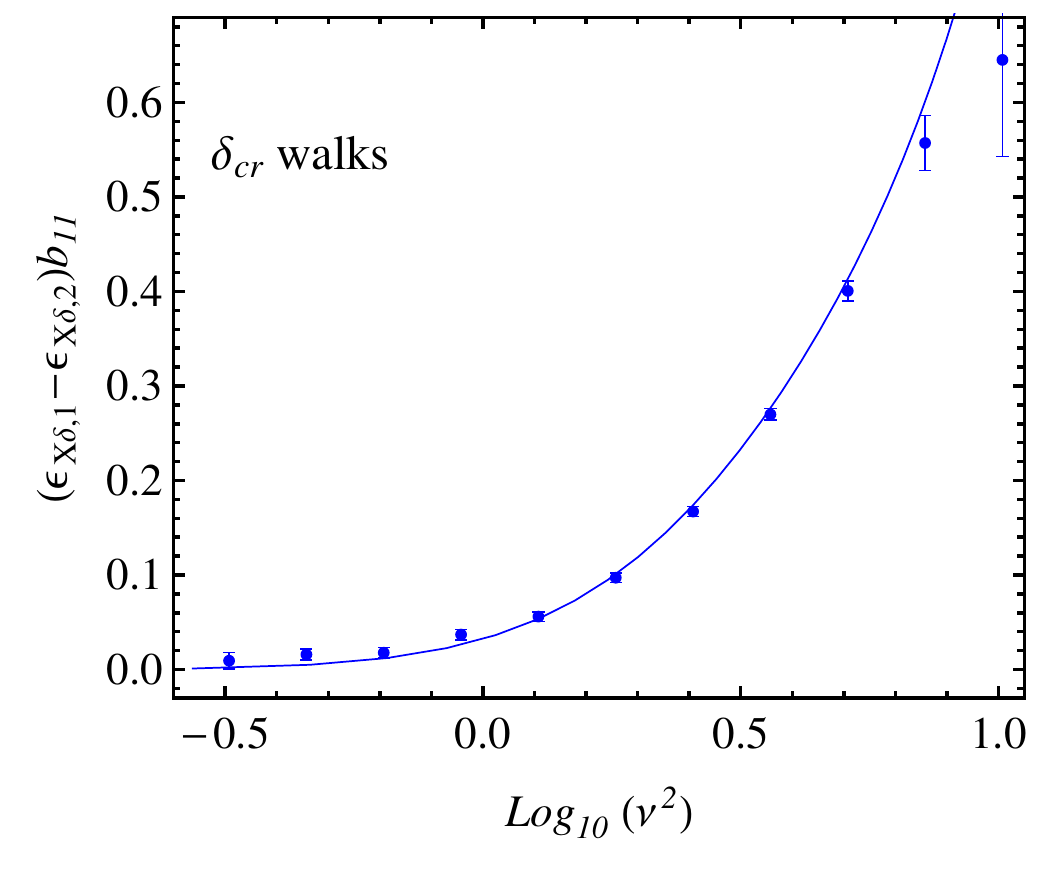}
 \caption{Same as previous Figure, but now $\widehat{b}_{11}$ has been multiplied by the appropriate $\epsilon_{Y\delta}$ coefficient (the determinant of the $2\times 2$ matrix in equation~(\ref{2x2}).  Top panel shows \eqn{eq:b11_hat}, which uses $\Delta$ and $X$ on the same smoothing scale $R$; the two sets of symbols are for two choices of $R$.  Bottom panel uses $X$ on the two different scales. Smooth curves show the associated predictions.  
}
 \label{fig:b11x}
\end{figure}

Figure~\ref{fig:b11} illustrates this for a number of different combinations of $Y$.  The symbols show four different estimates of $b_{11}$ for the walks with constant barrier $\delta_c$ and the curve shows equation~(\ref{diffb1s}).  Measurements in blue were obtained from \eqn{eq:b11_hat} at the same smoothing scales that were used in \fig{fig:b1s}. The other sets of symbols (red circles and squares) show the result of estimating the bias from a single field but at two different smoothing scales. Clearly, some estimates are noisier than others, but over all, there is general agreement with the prediction.  

Most of the measurement error comes from the smallness of the $\epsilon_{Y\delta}$ factors.  E.g., estimates from $\Delta$ at two (not very different) scales do not provide good estimates of $b_{11}$, because the differences between the two sets of measurements are small (c.f. \fig{fig:b1s}).  To study this further, we rescaled the measurements by the determinant of the matrix in equation~(\ref{2x2}) (for equation~\ref{eq:b11_hat}, this factor is $[\epsilon_{\Delta\delta}-\epsilon_{X\delta}]$).  Figure~\ref{fig:b11x} shows the results.  The top panel shows the estimator of equation~(\ref{eq:b11_hat}), which uses $\Delta$ and $X$ on the same scale $R$ (the two sets of symbols are for two choices of $R$); the bottom panel uses $X$ on two different scales for which the multiplicative factor is $[\epsilon_{X\delta}(R_1)-\epsilon_{X\delta}(R_2)]$; and the smooth curves show the associated predictions.  Most of the noise in the previous Figure is gone, so it is easy to see the predictions and measurements are in excellent agreement.  

\begin{figure}
 \centering
 \includegraphics[width=0.9\hsize]{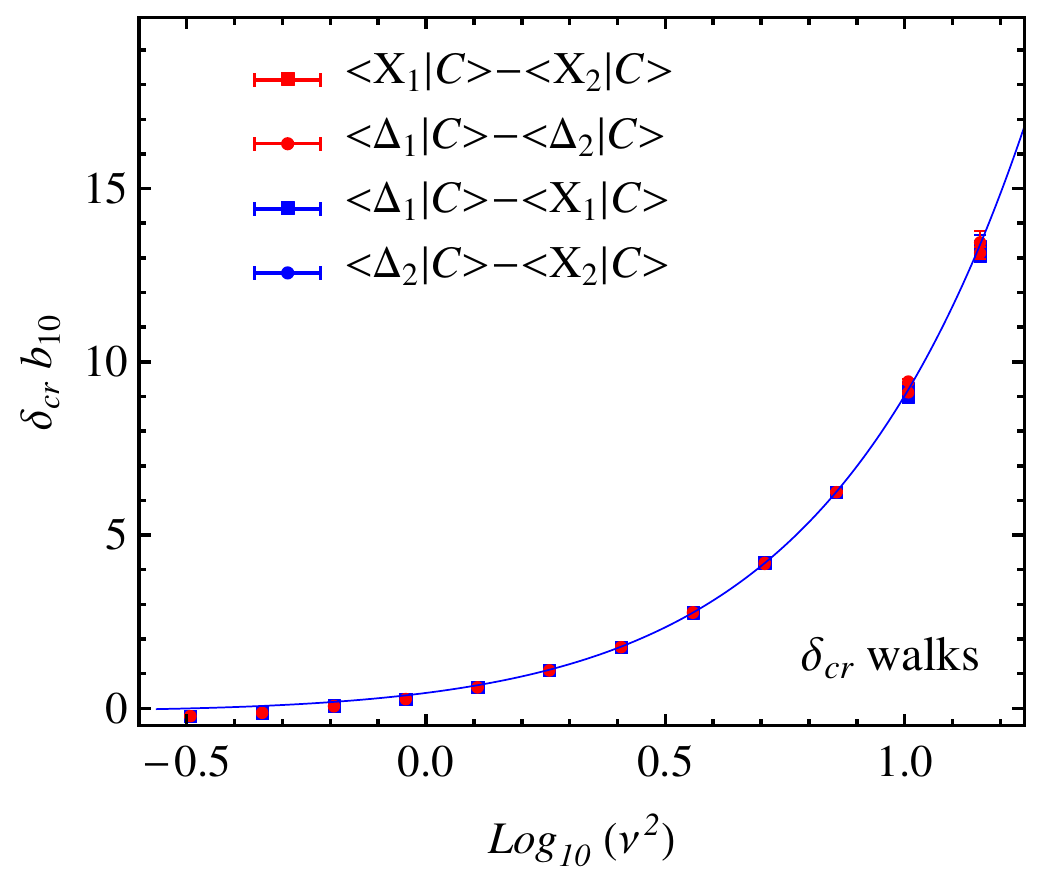}
 \caption{Comparison of $b_{10}$ with a number of estimates which were obtained from combining the Hermite polynomial weighted estimate of $b_1$ shown in Figure~\ref{fig:b1s} with the estimates of $b_{11}$ based on equation~(\ref{diffb1s}), for the different choices of $Y$ shown in the previous Figure (overdensity or its derivative measured on the same or different scales).}
 \label{fig:b10}
\end{figure}

The next step is to use the estimates of $b_{11}$ shown in Figure~\ref{fig:b1s} to correct the $b_1$ measurements shown in Figure~\ref{fig:b1s} (using equation~\ref{byr}) and so obtain $b_{10}$ without any assumptions about $\delta_c(s)$.  
Since our estimator for $b_{10}$ involves subtracting our estimators for $b_1$ and $b_{11}$, it can be written as a weighted sum of the two profiles $\widehat{b}_\Delta$ and $\widehat{b}_X$ that we used to construct $b_{11}$:
\begin{equation}
 \widehat{b}_{10}(s) = \frac{\epsilon_{X\delta}}{\epsilon_{X\delta}-\epsilon_{\Delta\delta}}\,\widehat{b}_\Delta
 + \frac{\epsilon_{\Delta\delta}}{\epsilon_{\Delta\delta}-\epsilon_{X\delta}}\,\widehat{b}_X.
\end{equation}
Figure~\ref{fig:b10} shows the results.  The agreement between the four estimators we have tried is remarkable.  (Measurements of the full $b_1$ have greater signal-to-noise; estimating the scale independent part $b_{10}$ requires the use of a different field, in our case $X$, and this may reduce the signal-to-noise of the measurement.)  We conclude that we are now able to estimate $b_{10}$ with no prior knowledge of $\delta_c$.  

\begin{figure}
 \centering
 \includegraphics[width=0.9\hsize]{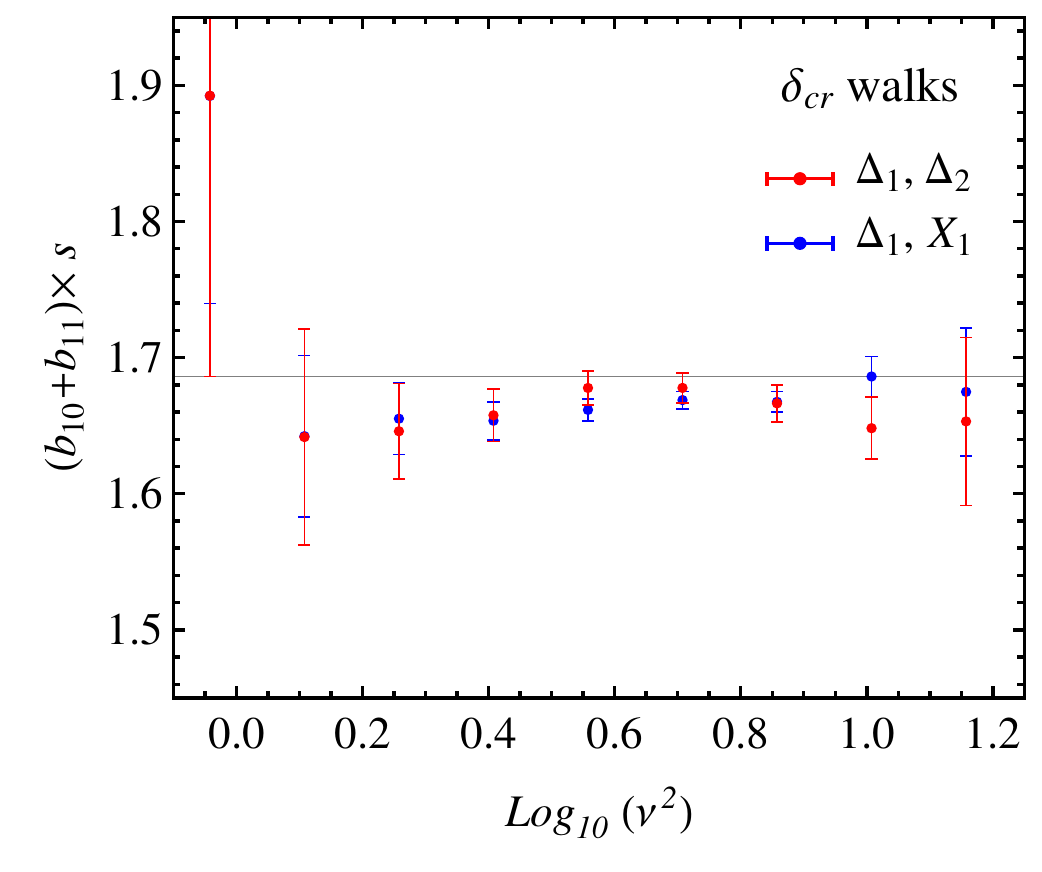}
 \caption{Estimate of $\delta_c$ from summing our previous estimates of $b_{10}$ and $b_{11}$ for the same choices of $Y$ shown in the previous Figure (overdensity or its derivative measured on the same or different scales).  Horizontal line shows the correct value; symbols show our various estimates.}
 \label{fig:bias2dc}
\end{figure}

We have made the point that our estimates of $b_{10}$ and $b_{11}$ were made without knowledge of $\delta_c(s)$.  Therefore, it is interesting to simply add these estimates, as equation~(\ref{b11fromb10}) indicates that they should sum to give $\delta_c(s)/s$.  In terms of the two profiles, our estimator reads 
\begin{equation}
 \widehat{\delta_c}(s) = s\left(
 \frac{1-\epsilon_{X\delta}}{\epsilon_{\Delta\delta}-\epsilon_{X\delta}} \widehat{b}_\Delta
 + \frac{1-\epsilon_{\Delta\delta}}{\epsilon_{X\delta}-\epsilon_{\Delta\delta}}\widehat{b}_X\right).
\end{equation}
Figure~\ref{fig:bias2dc} shows that this procedure works quite well.  In the next section we show that it also works when $\delta_c$ is stochastic, with a mean value which is scale dependent ($\delta_c(s)$ is a function of $s$).  We conclude that we are able to infer the value of $\delta_c(s)$, and potentially its dependence on $s$, from measurements of the surrounding density field, without any a priori information about the physics of halo formation.  

It is worth emphasizing how remarkable this is:  to date, it has been thought that one must know something about the physics of halo formation -- $\delta_c(s)$ -- to correctly predict halo abundances and bias.  We have shown that one can turn the argument around:  One can obtain interesting constraints on this physics from measurements of halo bias.  

\begin{figure}
 \centering
 \includegraphics[width=0.9\hsize]{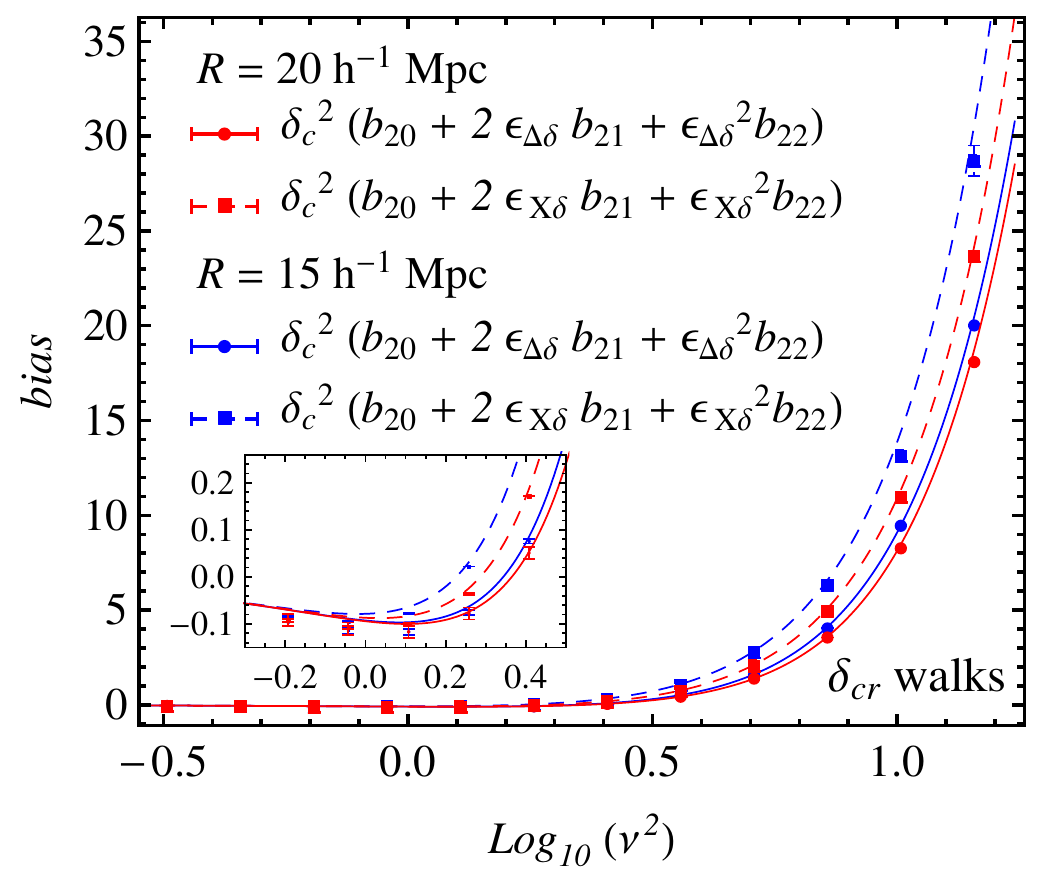}
 \caption{Estimate of $b_2$ (equation~\ref{eq:b2_hat}) for the choices of $Y$ indicated (overdensity or its derivative measured on the same or different scales).  Curves show the predicted (scale-dependent) value.}
 \label{fig:b2}
\end{figure}

We have also tested our methodology for estimating the second order bias parameters.  I.e., we first set 
\be
 \widehat{b}_{2Y} = \frac{\langle Y^2\rangle}{\langle Y\delta_p\rangle^2}
 \frac{1}{N}\sum_{\alpha=1}^N \left[\frac{Y_\alpha^2}{\langle Y_\alpha^2\rangle} - 1\right]
 \label{eq:b2_hat}
\ee
for three choices of $Y$, and then use equation~(\ref{3x3}) to estimate $b_{20}$, $b_{21}$ and $b_{22}$.  Figure~\ref{fig:b2} shows that our various estimators of $b_2$ depend on scale in the expected way:  scale-independent coefficients are multiplied by scale-dependent factors $\epsilon_{Y\delta}$ -- it is only these scale-dependent factors which depend on the choice of $Y$.

\begin{figure}
 \centering
 \includegraphics[width=0.9\hsize]{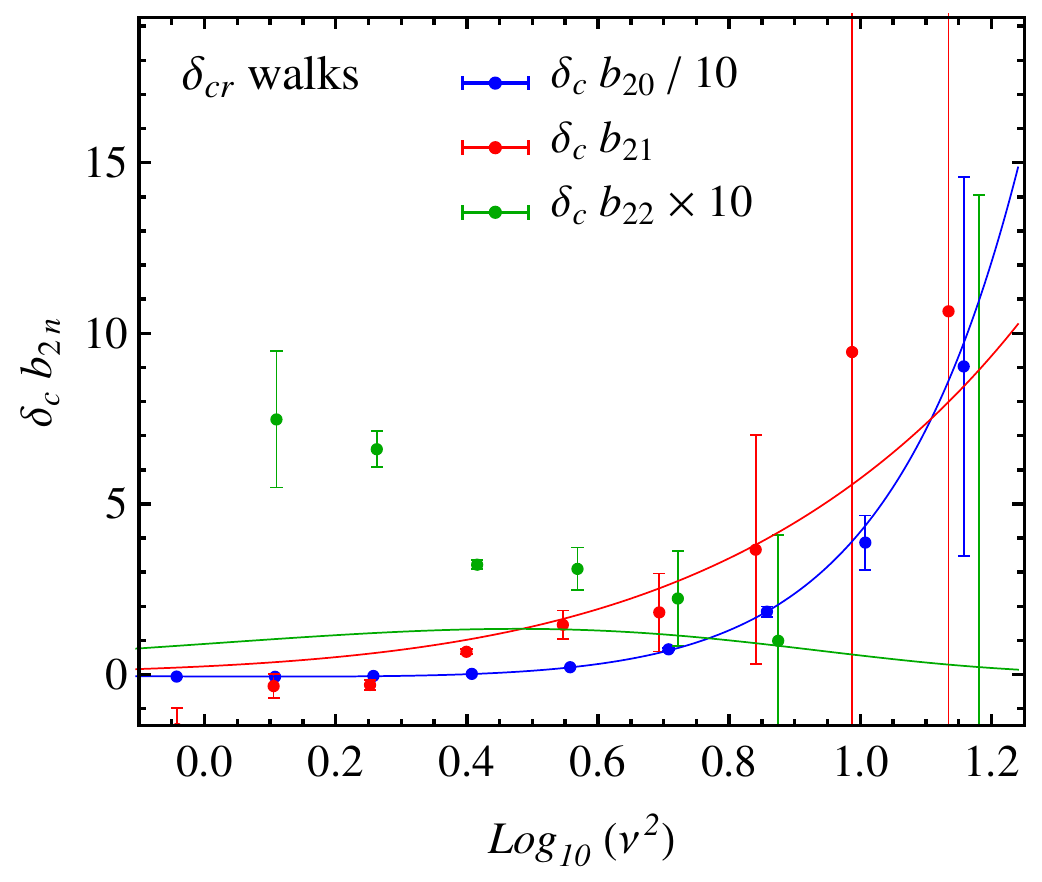}
 \caption{Estimates of the scale-independent bias coefficients $b_{20}$, $b_{21}$ and $b_{22}$ from inserting the various estimates of $b_2$ shown in the previous Figure in equation~(\ref{3x3}).  Curves show the predicted values.  }
 \label{fig:b2n}
\end{figure}

Figure~\ref{fig:b2n} shows our estimates of the scale-independent coefficients, obtained from using the $\widehat{b}_2$ values shown in Figure~\ref{fig:b2} in equation~(\ref{3x3}). These show that we recover $b_{20}$ very well, $b_{21}$ less well, and $b_{22}$ not well at all.  Our experience with $b_{11}$ means this is not completely unexpected -- we are differencing similar numbers and then normalizing by a small number.  (Notice that $b_{21}$ is about $10\times$ smaller than $b_{20}$, and $b_{22}$ is $10\times$ smaller still.)  Presumably, multiplying by the determinant of the matrix in equation~(\ref{3x3}) would reduce some of this systematic, for the same reason that theory and measurement are in better agreement in Figure~\ref{fig:b11x} than in Figure~\ref{fig:b11}.  

Despite the disagreement for $b_{22}$, it is worth pausing to appreciate what Figure~\ref{fig:b2n} shows.  The value of $b_{20}$ is usually estimated from measurements on large scales.  We have been able to estimate it remarkably well from measurements on much smaller scales -- scales which are almost nonlinear.  And $b_{21}$ has never been measured before.  

\begin{figure}
 \centering
 \includegraphics[width=0.9\hsize]{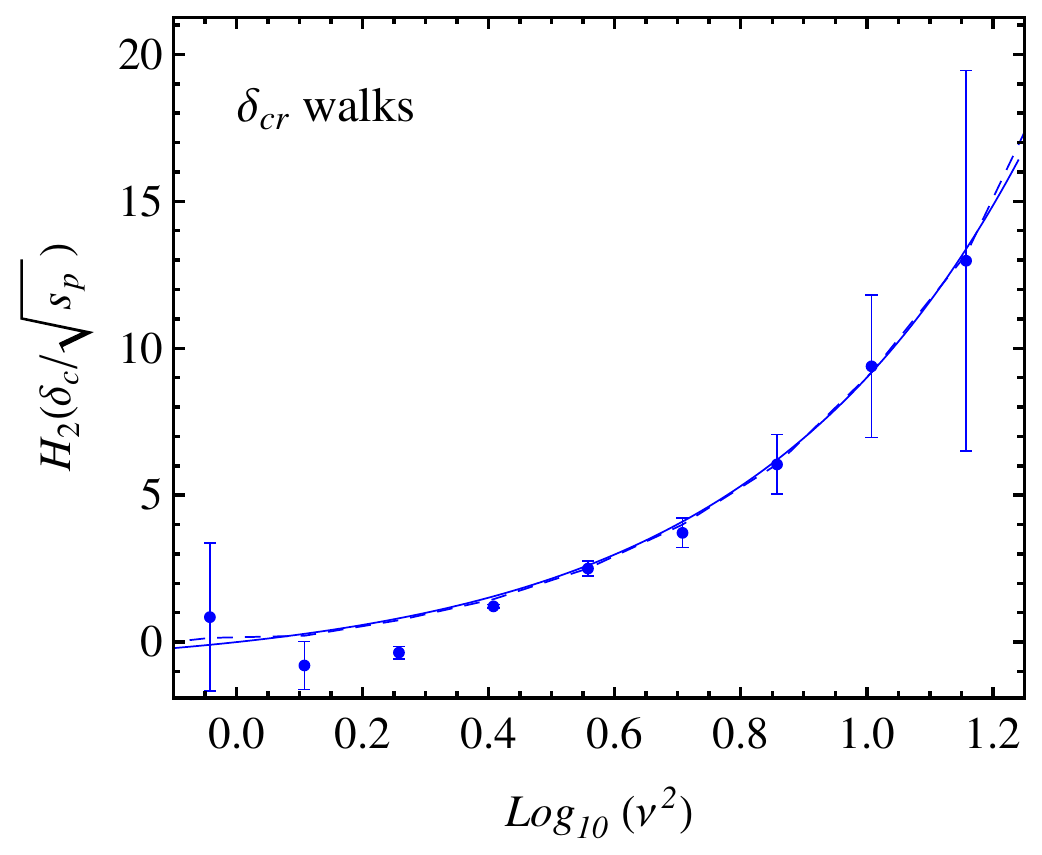}
 \caption{Test of consistency relation (equation~\ref{b2todc}).  Symbols show the result of inserting the $b_{20}$, $b_{21}$ and $b_{22}$ estimates shown in the previous Figure into the left hand side of equation~(\ref{b2todc}).  Solid curve shows the predicted value; dashed curve shows the result of inserting the estimate of $\delta_c$ shown in Figure~\ref{fig:bias2dc} into the right hand side of equation~\ref{b2todc}.  The agreement shows that we recover the same $\delta_c$ values from our measurements of $b_2$ that we did from $b_1$. }
 \label{fig:b2todc}
\end{figure}

While the smallness of $b_{21}$ and $b_{22}$ has made estimating them difficult, it also means that we should be able to make a relatively clean test of the consistency relation, equation~(\ref{b2todc}).  The symbols in Figure~\ref{fig:b2todc} show the result of inserting the estimates shown in Figure~\ref{fig:b2n} into the left hand side of equation~(\ref{b2todc}) (and multiplying by $s_p$).  The solid curve shows the predicted value -- the right hand side of equation~(\ref{b2todc}) (times $s_p$).  

One might argue that we were only able to show the theory curve because we knew the value of $\delta_c$ to begin with.  We could, of course, have assumed we did not know it, and then fit the measurements to $H_2(a\nu)$, to see if the parameter $a$ is constant or not.  We leave such tests for future work.  For the present purposes, we think it is sufficient that the agreement between theory and measurement in Figure~\ref{fig:b2todc} indicates that our methodology has indeed recovered the correct value of $\delta_c$, and the fact that it is independent of scale.  

As another consistency check, the dashed curve shows the result of using the estimate of $\delta_c$ from Figure~\ref{fig:bias2dc} -- that based on the linear bias factors $b_{10}$ and $b_{11}$  -- in the right hand side of equation~(\ref{b2todc}).  The agreement is very good, again suggesting that our methodology works well.  

Before moving on, we note again that these consistency checks are unprecedented.  Never before has the nonlinear bias factor $b_2$, measured on nonlinear scales, been used to estimate $\delta_c$.  Perhaps more remarkably, our estimate of $\delta_c$ is made from {\em linear} combinations of nonlinear bias factors measured on nonlinear scales.

\section{Stochasticity}
\label{subsec:qwalks}
The previous section showed that one gets the same $k$-dependent bias by cross-correlating the protohaloes with any field $Y$ that is correlated with one of the variables which determine the protohalo constraints ${\cal C}_p$, if bias is defined (as is conventional) with respect to $\avg{\!Y\delta_p\!}$.  It also showed that this allows one to estimate the $k$-independent bias factor $b_{10}$ (and, by extension, all the bias factors $b_{nj}$ as well) with no prior assumptions about the value of $\delta_c(s)$.  

One might have wondered if this only works if the barrier is a deterministic function of $s$.  To show that it is more general, we now consider a model in which first crossing is determined by two variables $\delta$ and $q$, and their derivatives:
\begin{eqnarray}
 \delta(R_p) &\ge& \delc(1+ \sqrt{s_p}\,q(R_p)/q_c) \notag\\
 \delta(R_p+\Delta R) &\le& \delc[1+ \sqrt{s_{p+\Delta}}\,q(R_p+\Delta R)/q_c].
 \label{dcq}
\end{eqnarray}
Here, $q$ may or may not have the same statistical properties as $\delta$.  E.g., even if both $\delta$ and $q$ are Gaussian, they may have different correlation properties.  The case in which $\delta$ is Gaussian but $q$ is not can be related to models in which both the overdensity and the traceless shear determine halo formation \cite[e.g.][]{st02}.  
 In such models $q$ is independent of $\delta$, and one may think of the conditions above as defining a model in which the critical overdensity for collapse varies stochastically \cite[][]{scs12,ms14}.  

This is a nice model to explore in the present context since the distribution $p(q^2)$ is chi-squared with 5 degrees of freedom,
\be
\label{eq:defq2}
 q^2 \equiv \frac{1}{5}\sum_{i=1}^5 \frac{g_i^2}{s}\,,
\ee 
 where the $g_i$ are Gaussian random variables with zero mean and variance $s$ (we will sometimes use the notation $q_5$ instead of $q$).  Therefore, unlike $\delta$ (which is Gaussian distributed), it has non-zero mean.  Thus, if one wished to ignore the stochasticity arising from the distribution of $q_5$ by replacing $q_5\to\avg{q_5}$ in equation~(\ref{dcq}), then we would have a problem in which $\delta$ must cross a deterministic barrier whose height depends on $s$.  The deterministic and constant barrier of the previous section corresponds to $q_c\to\infty$.  Since our more general model is more complex in both respects, we now wish to explore the consequences of the fact that ${\cal C}_p$ now depends on $\delta_p$ and $v_p$ (as in the previous section) as well as $q_p$ and its derivative.  

\begin{figure}
 \centering
 \includegraphics[width=.9\hsize]{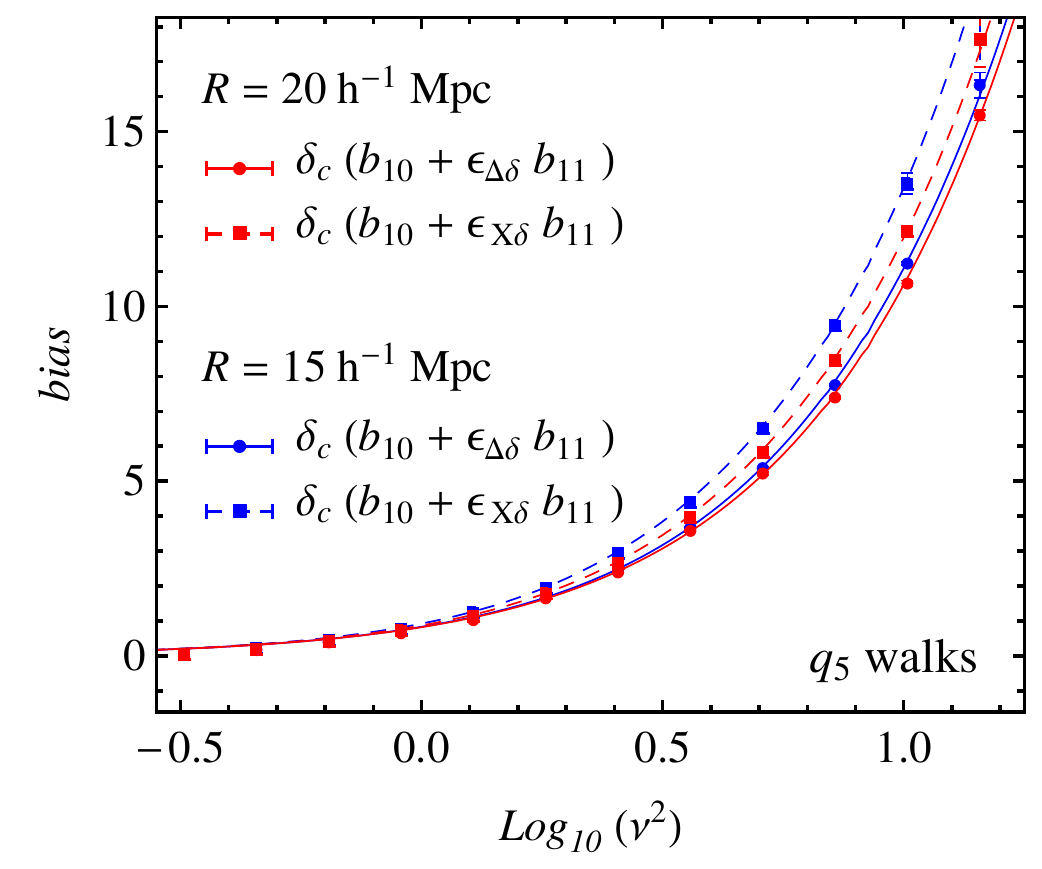}
 \caption{Same as \fig{fig:b1s} but for walks crossing a stochastic boundary (equation~\ref{dcq}) that has $q_c =6.25$.}
 \label{fig:b1s_q5}
\end{figure}

\subsection{Consistency relations for density bias}
For the same reasons as before, we expect this model to have bias factors $b_{10}$ and $b_{11}$ which depend on $\delta_c$ and $s$.  However, because the field $q$ also matters, we expect the values of $b_{10}$ and $b_{11}$ to depend on $q_c$ as well.  In particular, we expect equation~(\ref{b11fromb10}) to become 
\be
 b_{10} + b_{11} = \frac{\avg{\delta_{1\times}}}{s} = 
 \frac{\delta_c}{s}\,\left(1 +\sqrt{s} \frac{\avg{q_p|\mathcal{C}_p}}{q_c}\right) 
 \label{eq:bias2dcq}
\ee
where $\avg{\delta_{1\times}}\equiv \avg{\delta_p|\mathcal{C}_p}$ is the mean value of $\delta$ on the first crossing scale.  The final equality shows that this mean value is related to the mean value of $q$ at first crossing, which we might write in more suggestive notation as  $\avg{q_p|\mathcal{C}_p}\equiv \avg{q_{1\times}}$.  More generally, we expect equation~(\ref{bn->dc}) to become 
\be
 \sum_{j=0}^n {n\choose j}\, b_{nj} = s^{-n/2}\,\Bigl\langle H_n(\nu_{1\times})\Bigr\rangle.
 \label{eq:cons_q}
\ee
\citep{cphs17}, where we have set $\nu_{1\times}\equiv\delta_{1\times}/\sqrt{s}$.

On the other hand, because $q$ is independent of $\delta$, we do not expect $q$ to contribute scale (or $k$-) dependence to correlations with $\delta$.  Therefore all the technology of the previous section should go through unchanged:  we should be able to estimate the scale independent $b_{10}$ from the scale dependent $b_1$ using any combination of cross-correlations of $\Delta$ and its derivatives around the first crossing scales (i.e. centered on the protohalo positions).

\begin{figure}
 \centering
 \includegraphics[width=0.9\hsize]{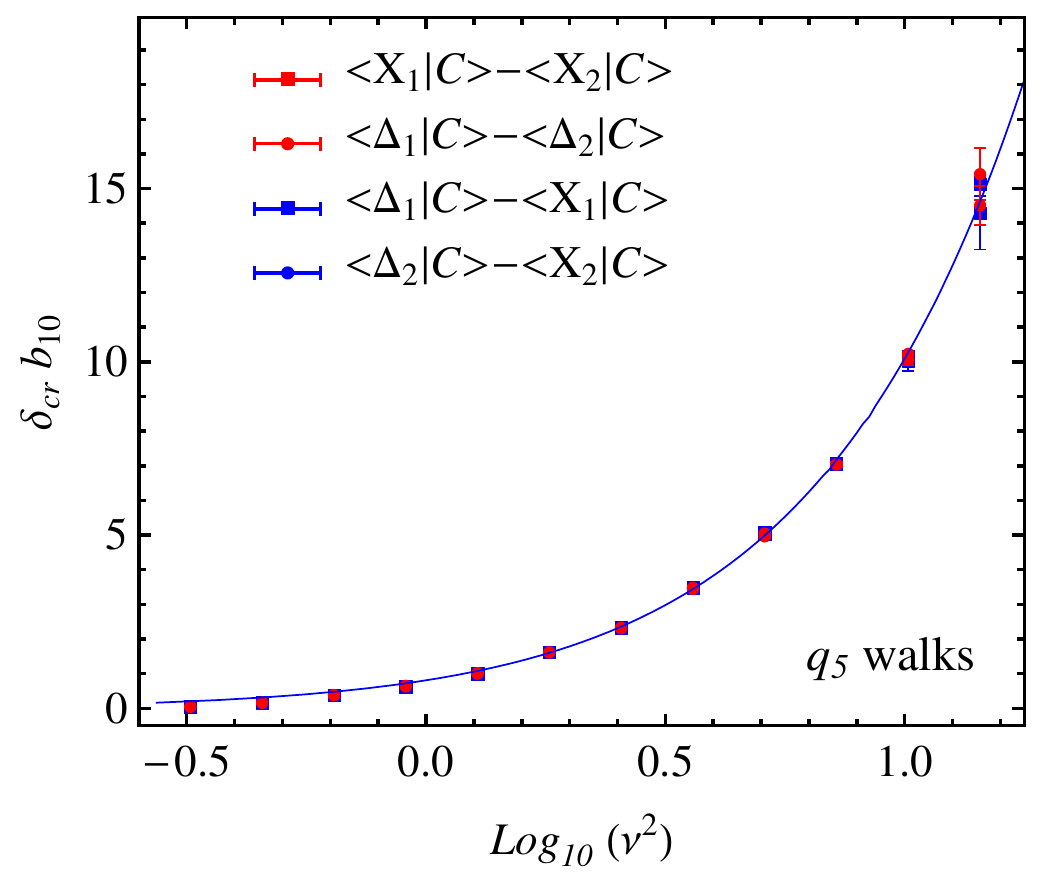}
 \caption{Same as Figure~\ref{fig:b10}, but now for the case in which $q_c=6.25$. }
 \label{fig:b10q}
\end{figure}

Figure~\ref{fig:b1s_q5} shows $b_1$ for this stochastic barrier model in which $q_c=6.25$.  The agreement between the theoretical curves and the measurements is as good as in \fig{fig:b1s} (where the barrier was deterministic).  Figure~\ref{fig:b10q} shows estimates of  $b_{10}$ obtained analogously to the constant deterministic barrier case shown in Figure~\ref{fig:b10}.  The agreement between our various estimators is again excellent.  This is a nontrivial extension of the constant barrier one shown in Figure~\ref{fig:b10} since, in this case, not only is the barrier stochastic, but the fact that the mean of $q$ is non-zero makes the effective value of $\delta_c(s)$ increase with $s$.  Thus, we conclude that we are now able to estimate $b_{10}$ with no prior knowledge of $\delta_c$ (constant or not? stochastic or not? etc.).  

\begin{figure}
 \centering
 \includegraphics[width=0.9\hsize]{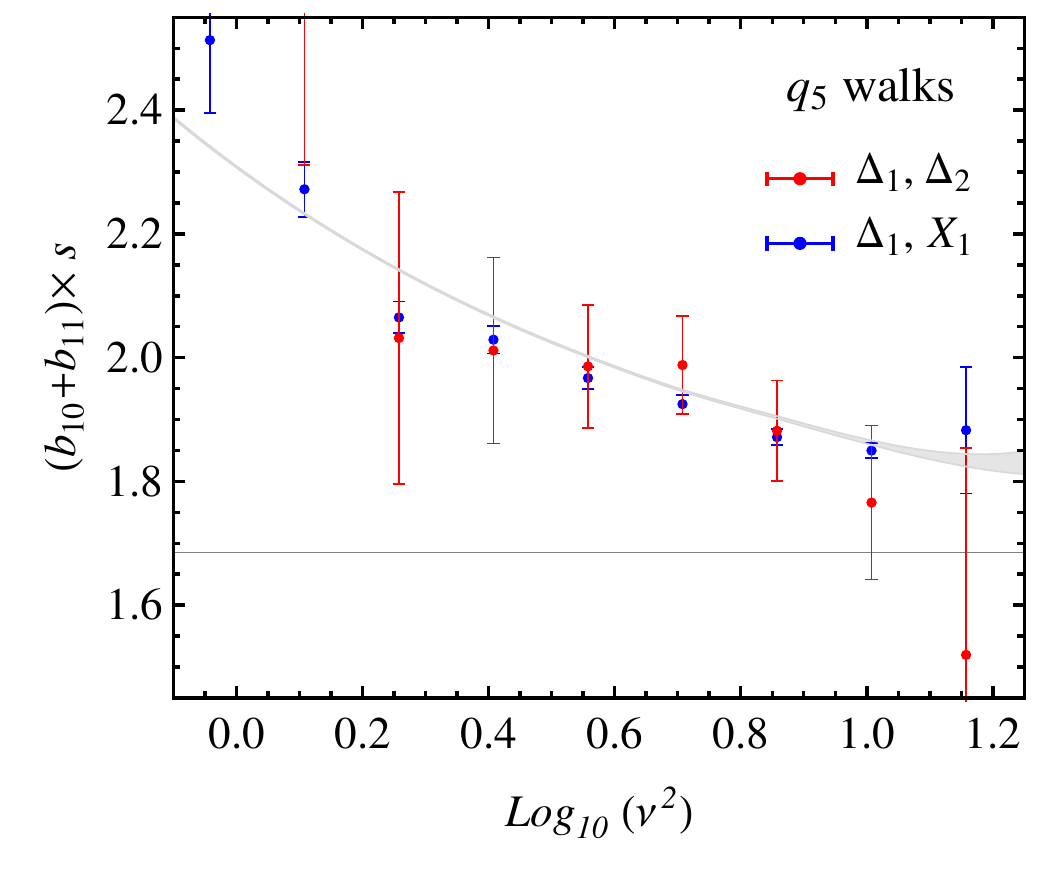}
 \caption{Same as Figure~\ref{fig:bias2dc}, but now for the case in which $q_c=6.25$.  In this model, the mean value of $\delta_c(s)$ increases with $s$ because of stochasticity (grey band).  Our methodology correctly reproduces this trend with no prior knowledge of either the stochasticity or the scale dependence.  }
 \label{fig:bias2dcq}
\end{figure}

Following the steps laid out in the previous section, we can also estimate $b_{11}$.  This works well, so we have not shown it.  Rather, we have combined it with the estimate of $b_{10}$ shown in Figure~\ref{fig:b10q} to estimate $\delta_c(s)$.  Figure~\ref{fig:bias2dcq} shows that this works very well:  in particular, our methodology is able to recover the (stochasticity induced) mass dependence of $\delta_c$ quite faithfully.  

\begin{figure}
 \centering
 \includegraphics[width=0.9\hsize]{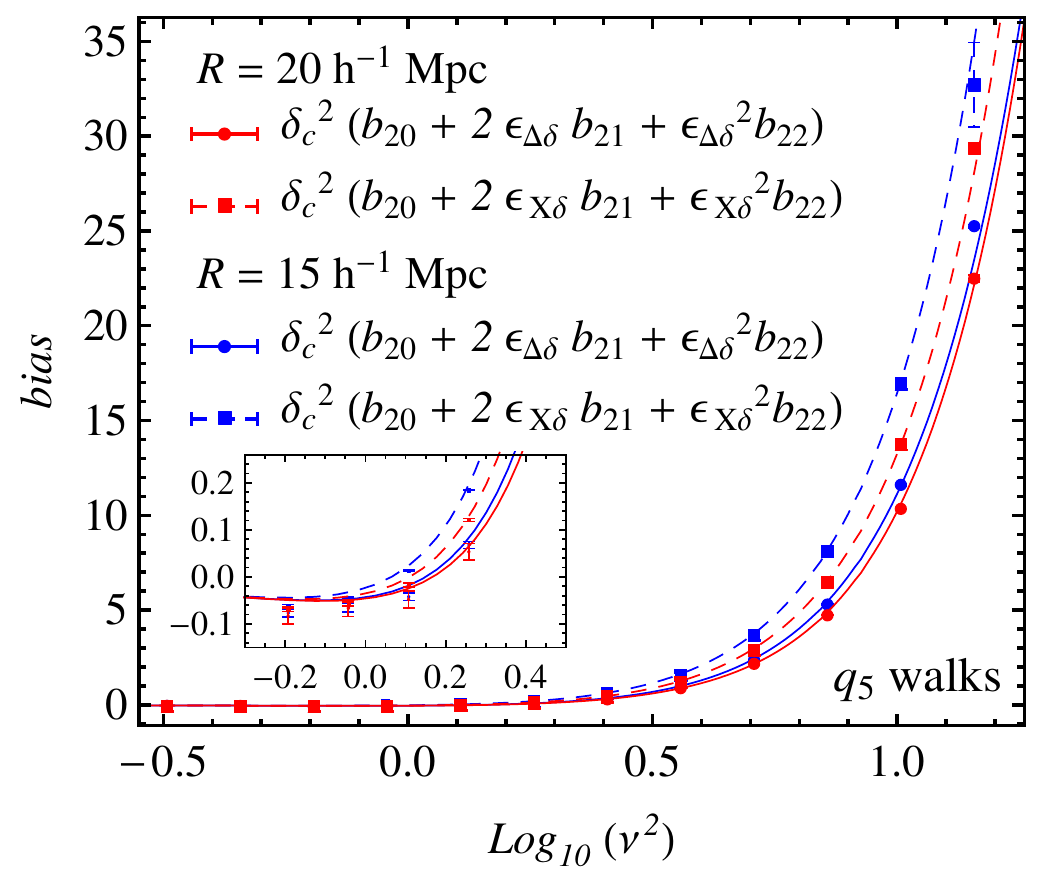}
 \caption{Same as Figure~\ref{fig:b2}, but now for the case in which $q_c=6.5$.}
 \label{fig:b2q}
\end{figure}

\begin{figure}
 \centering
 \includegraphics[width=0.9\hsize]{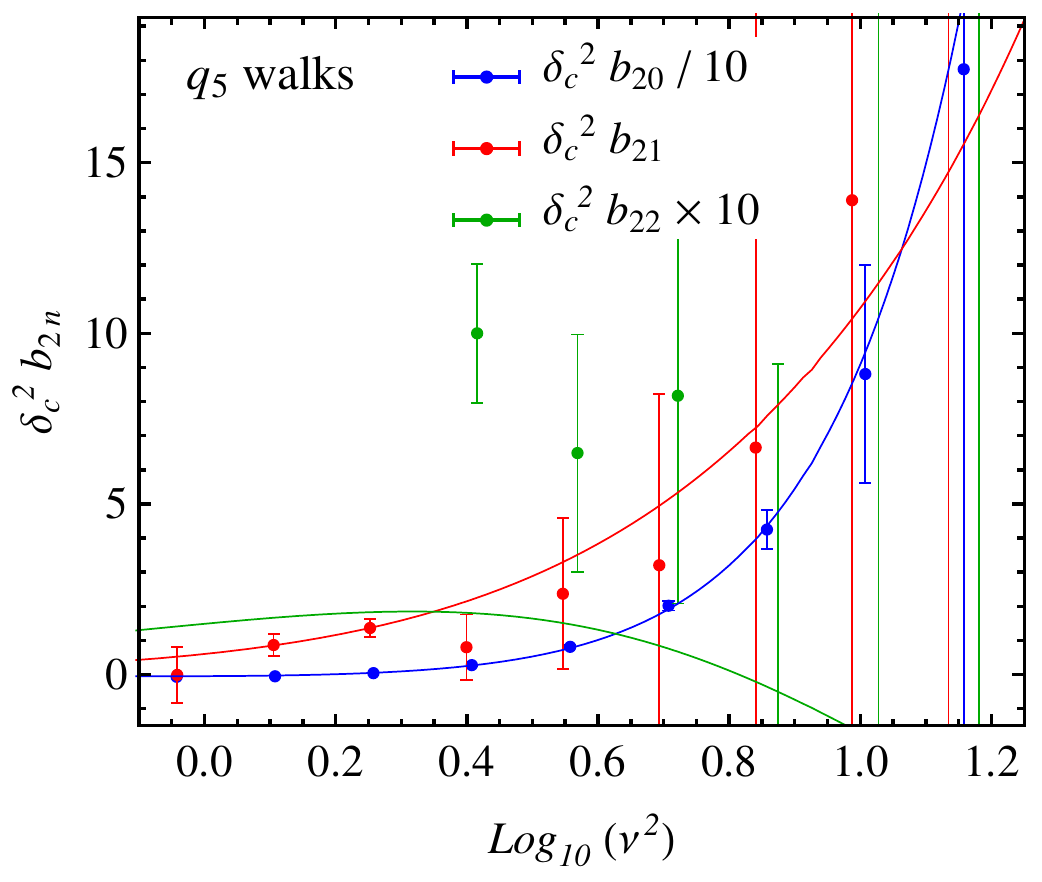}
 \caption{Same as Figure~\ref{fig:b2n}, but now for the case in which $q_c=6.5$.  I.e., estimates of the scale-independent bias coefficients $b_{20}$, $b_{21}$ and $b_{22}$ come from inserting the various estimates of $b_2$ shown in the previous Figure into equation~(\ref{3x3}).  Curves show the predicted values.  The estimates are noisier than those in Figure~\ref{fig:b2n} because of the stochasticity coming from the variable $q$.}
 \label{fig:b2nq}
\end{figure}

\begin{figure}
 \centering
 \includegraphics[width=0.9\hsize]{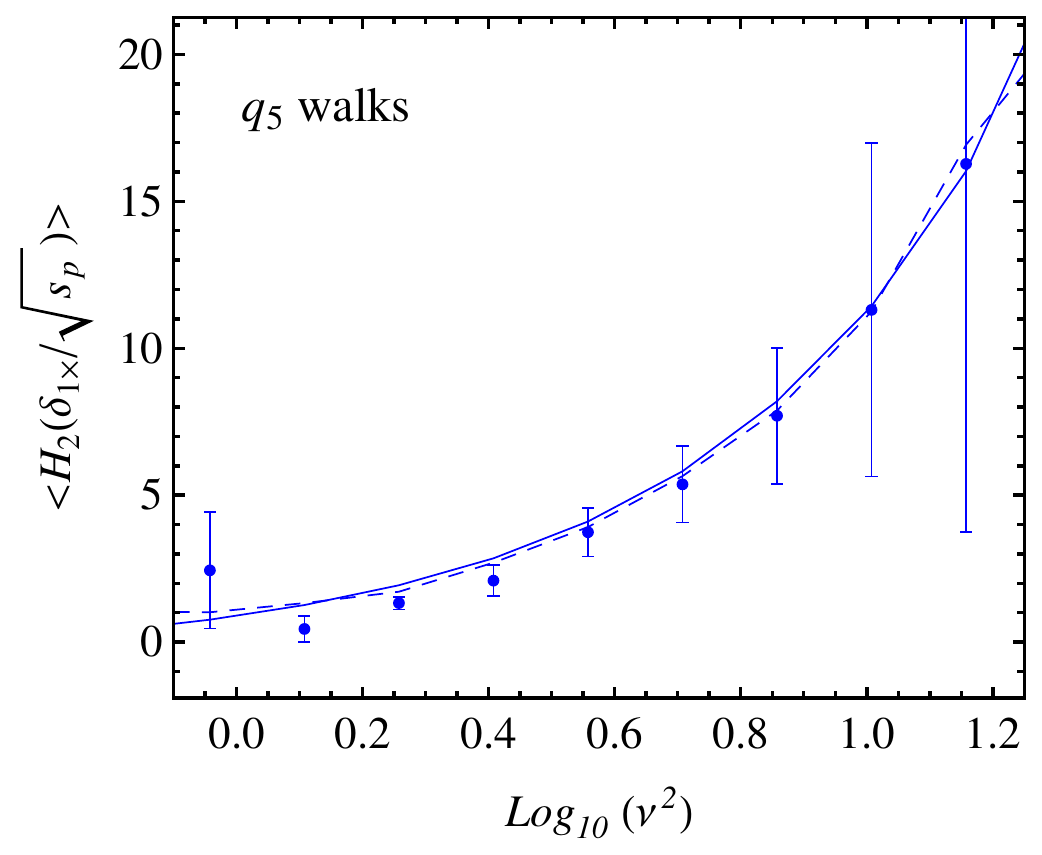}
 \caption{Test of consistency relation (equation~\ref{b2todc}):  Symbols show the result of inserting the $b_{20}$, $b_{21}$ and $b_{22}$ estimates shown in the previous Figure into the right hand side of equation~(\ref{b2todc}).  Solid curve shows the predicted value.  Dashed curve shows $H_2\left(\langle\delta_{1\times}\rangle/\sqrt{s_p}\right)$, with $\langle\delta_{1\times}\rangle$ given by Figure~\ref{fig:bias2dcq}. }
 \label{fig:b2q->dcq}
\end{figure}

Finally, the higher order bias coefficients $b_n$ can also be measured in the same way as before, by averaging higher-order Hermite polynomials centered on the constrained regions.  Figures~\ref{fig:b2q}--\ref{fig:b2q->dcq} respectively show the scale dependent estimate of $b_2$, the scale-independent bias coefficients $b_{20}$, $b_{21}$ and $b_{22}$ from inserting the various estimates of $b_2$ shown in the previous Figure into equation~(\ref{3x3}), and the consistency relation (equation~\ref{b2todc}).  In all cases, the methodology works as well here as it did for the constant deterministic barrier of the previous section -- although some of the estimates (e.g. $b_{22}$) are slightly noisier because of the stochasticity.  

\subsection{Signature of stochasticity}
In stochastic barrier models, the value of $\delta$ at first crossing on scale $s$ is not deterministic, but has a distribution $p(\delta_{1\times}|s)$.  Equation~(\ref{eq:cons_q}) with $n=1$ estimates the mean of this distribution, $\langle\delta_{1\times}|s\rangle$.  Figure~\ref{fig:bias2dcq} shows that we were able to correctly reconstruct the fact that it depends on $s$, even if this was not known a priori.  When $n=2$, then equation~(\ref{eq:cons_q}) can be combined with $\langle\delta_{1\times}|s\rangle$ to yield an expression for the variance around the mean in terms of the first and second order bias factors.  We will write the variance in two ways.  

First, we note that 
\be
 {\rm Var}(\delta_{1\times})/s
   = \avg{H_2(\nu_{1\times})} - H_2(\avg{\nu_{1\times}}).
 \label{eq:vard1x}
\ee 
Since non-zero variance is a signature of stochasticity, the expression above shows that the difference between $\avg{H_2(\nu_{1\times})}$ and $H_2(\avg{\nu_{1\times}})$ is a direct measure of stochasticity.  The solid and dashed curves in Figure~\ref{fig:b2q->dcq} show that these two quantities are actually rather similar -- evidently, the stochasticity in this particular model is small, for reasons we quantify in the next subsection.  

Using the consistency relations to write the Hermite polynomials above in terms of the $b_{ij}$ leads to our second suggestive expression for the variance:  
\be
 \frac{{\rm Var}(\delta_{1\times}|s)}{s}
   = 1 + s\, (b_{20}-b_{10}^2 + 2b_{20}-2b_{10}b_{11} + b_{22}-b_{11}^2).
 \label{varbij}
\ee 
This shows explicitly how measurements of the first and second order bias factors yield information about the rms stochasticity.  Of course, this argument can be generalized to the statement that measurements of $b_n$ constrain the $n$th order moments of $p(\delta_{1\times}|s)$.  We exploit this fact elsewhere.

\subsection{Two types of consistency relations for shear bias}\label{subsec:c2}
We noted above that, although the stochasticity provided by $q$ does affect the amplitude of scale dependent bias (by modifying the values of $b_{10}$ and $b_{11}$; one can see this by comparing the predicted bias at the highest $\nu$ shown in Figures~\ref{fig:b10} and~\ref{fig:b10q}), it does not contribute any new scale dependence.  However, we do expect to get entirely new bias factors, $c_{n}$, say, which arise because ${\cal C}_p$ depends on $q_p$ and its derivative (c.f. equation~\ref{dcq}).  Such terms are usually referred to as `nonlocal' bias coefficients \citep{css13,scs12,baldauf12}.  We expect these to be scale dependent and to satisfy their own consistency relations in terms of scale independent coefficients $c_{nj}$ \citep{cs13,cphs17}. Since $q$ is fundamentally quadratic with respect to Gaussian variables (equation~\ref{eq:defq2}), we expect these coefficients to only occur for even $n$. E.g., when $n=2$, we expect to be able to estimate $c_{20}$, $c_{21}$ and $c_{22}$ from cross-correlations with the shear field, without prior knowledge of $q_c$ and/or stochasticity; and we expect cross-correlations of all quantities which correlate with $q$ to furnish estimates of $c_2$.  We also expect all this to generalize to higher order nonlocal bias coefficients $c_{n}$.  

We consider the issue of consistency relations first.  The orthogonal polynomials associated with $\chi^2$-distributed variables are modified Laguerre polynomials (e.g. $L_1^{(\alpha)}(x) = 1+\alpha - x$).  In the Appendix, we show that consistency relations between the shear bias factors $c_{nj}$ are similar to those for the density provided one works with these orthogonal polynomials.  For example, in close analogy with \eqn{eq:cons_q} for $n=2$, 
\be
\label{eq:consrelq2}
c_{20} + 2 c_{21} + c_{22} = - \avg{L_1^{(3/2)}(5q_{1\times}^2/2)} \,
\ee
(also see equations~\ref{eq:c20App}-\ref{eq:c21App}).  Similarly, the coefficients $c_{nj}$ will sum to $L_{k}^{(3/2)}(5q_{1\times}^2/2s)$, where $n=2k$.  

The similarity in spirit between density and shear bias factors goes deeper.  In the previous section, we made the point that the density bias factors $b_{nj}$ can be rearranged to describe the moments of $p(\delta_{1\times}|s)$ (e.g. equation~\ref{eq:vard1x}).  For similar reasons, the shear bias factors $c_{nj}$ can be rearranged to describe the moments of $p(q_{1\times}|s)$.  While it is natural to expect this deeper connection, there is another type of consistency which is, perhaps, more surprising.  These arise because 
\be
\label{eq:q1x}
 \langle\delta_{1\times}^n\rangle = \langle\delta_c^n\,(1 + \sqrt{s}\,q_{1\times}/q_c)^n\rangle,
\ee
so moments of $\delta_{1\times}$ can be written as linear combinations of the moments of $q_{1\times}$ and vice versa.  Hence, for example, 
\begin{align}
\label{eq:H2stoc}
 \avg{H_2(\nu_{1\times})} 
    &= (\delta_c^2/s)\, \avg{1+  s (q_{1\times}/q_c)^2 + 2\sqrt{s}\, q_{1\times}/q_c} - 1 \nonumber\\
    &= H_2(\avg{\nu_{1\times}}) + (\delta_c/q_c)^2\, {\rm Var}(q_{1\times}),
\end{align}
where Var$(q_{1\times}) = \avg{q_{1\times}^2} - \avg{q_{1\times}}^2$.  Comparison with equation~(\ref{eq:vard1x}) shows that 
\be
 \frac{{\rm Var}(\nu_{1\times})}{\delta_c^2} = \frac{{\rm Var}(q_{1\times})}{q_c^2}.
\ee
Note that the difference between $\avg{H_2(\nu_{1\times})}$ and $H_2(\avg{\nu_{1\times}})$ equals 
$(\delta_c/q_c)^2$ times the variance of $q_{1\times}$.  In our model, both these factors are substantially smaller than $1$, which is why the solid and dashed curves in Figure~\ref{fig:b2q->dcq} are so similar.  

In addition, from \eqn{eq:q1x} and the first line in \eqn{eq:H2stoc} we have
\begin{align} 
\avg{L_1^{(3/2)}(5q_{1\times}^2/2s)} &= \frac{5q_c^2}{2\delta_c^2}\, \bigg[1 + \avg{H_2(\nu_{1\times})} - 2\nu \avg{H_1(\nu_{1\times})}\notag\\
&\ph{5q_c^2(1+\nu^2)}
 + \nu^2 - (\delta_c/q_c)^2\bigg].
 \label{eq:cons_c2}
\end{align}
This is a remarkable expression; it shows that the density bias factors $b_{nj}$ and those for the shear $c_{nj}$ are related, even though the density $\delta$ and the shear $q$ are independent.  As a result of these relations, one can get information about the nonlocal bias factors $c_{nj}$ simply by taking appropriate combinations of cross correlations with the density rather than the shear field.  Since the former are much easier to measure, this represents a substantial simplification.  Perhaps more importantly, the expression above shows that measurements of $b_{nj}$ constrain the parameter space of allowed $c_{nj}$ values.

\subsection{Estimators and tests of consistency relations}
We turn now to direct estimates of the shear bias factors.  By analogy with the deterministic barrier case, we set 
\begin{align}
 \widehat{c}_{2,q_5}(s) = -\frac{1}{r^2} \frac{1}{N}
                         \sum_{\alpha=1}^N L_1^{(3/2)}(5Q^2_{5,\alpha}/2) 
\label{eq:c2_hat}
\end{align}
where the sum runs over all the $N$ walks that cross in a given narrow bin of width $\Delta s$ centered on $s$, $Q_5$ denotes the value of the shear field on some smoothing scale that is larger than the one associated with first crossing and $r \equiv S_{0\times/}\sqrt{s S}$.\footnote{We defined $S_{j\times} \equiv (2\pi)^{-3}\int \mathrm{d}^3k\,k^{2j} P_{\delta\delta} W_{R_p} W_{R_q}$ and $S_{j} \equiv (2\pi)^{-3}\int \mathrm{d^3}k\,k^{2j} P_{\delta\delta} W_{R_q}^2$.}  
Also by analogy with $\delta$, we expect $\widehat{c}_{2,q_5}(s)$ to depend on a piece, $c_{20}$, which dominates on large scales, and two others, $c_{21}$ and $c_{22}$, which yield scale dependence.  

\begin{figure}
 \centering
 \includegraphics[width=0.9\hsize]{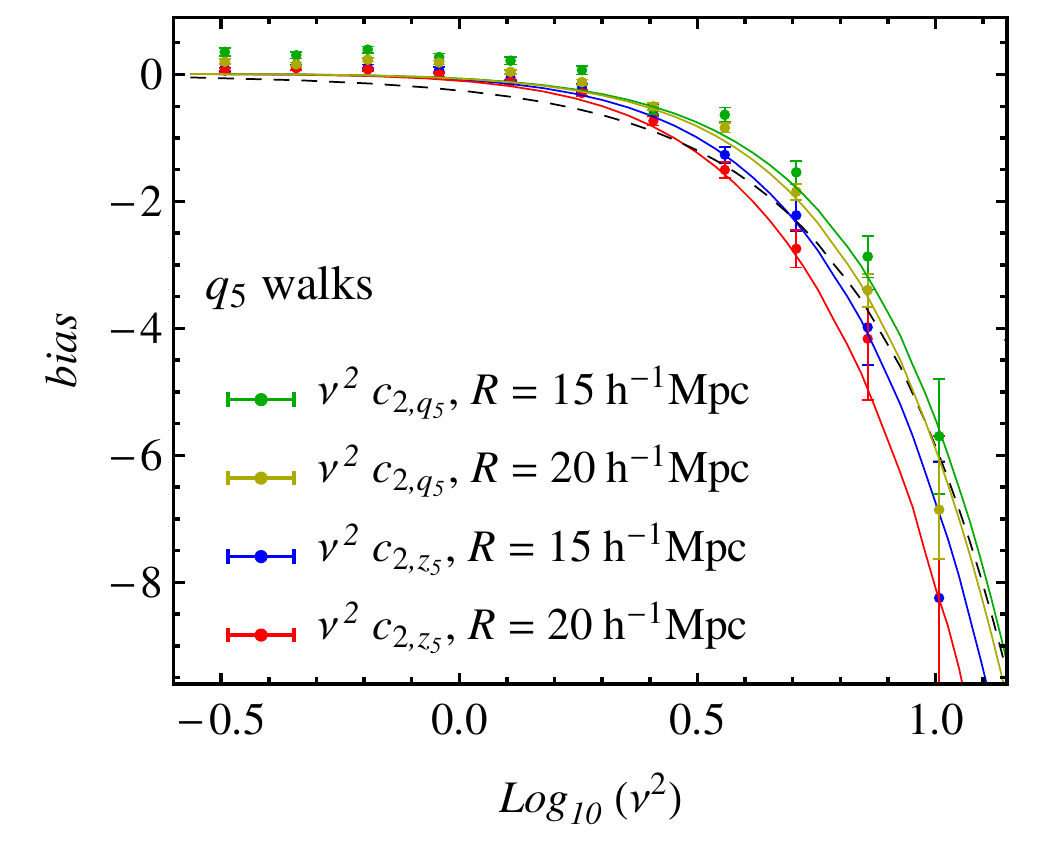}
 \caption{Same as Figure~\ref{fig:b1s}, but now for the case in which $q_c=6.25$ and we use cross-correlations (equation~\ref{eq:c2_hat}) to estimate $c_2$ rather than $b_1$.  Two sets of symbols are for cross-correlating with the shear on two scales, and another two are for the shape (equation~\ref{eq:r}) on two scales.  All the measurements are indistinguishable, and are in excellent agreement with the solid curve which shows equation~(\ref{eq:c20}).}
 \label{fig:c2s}
\end{figure}

Figure~\ref{fig:c2s} shows results for two different smoothing scales; for now, consider only the top two sets of (green and yellow) symbols.  In both cases, we show $\nu^2\,\widehat{c}_{2,q_5}(s)$, for ease of comparison with \cite{scs12}:  $\delta_c^2\,c_{2}^{\rm SCS} = 2\nu^2 c_{20}$.  First, notice that the signal is always negative.  As discussed by \cite{scs12}, this is a result of the model assumption that the shear, $q^2$, inhibits the formation of objects (because it adds to $\delta_c$ in equation~\ref{dcq}).  Second, it is hard to distinguish between the green and yellow symbols:  evidently, $c_{2,q_5}$ depends less strongly on scale than does $b_1$.  In the $R\to\infty$ limit, we expect the dominant contribution to be (see Appendix~\ref{App:A} and~\ref{app:EScon} for details) 
\begin{align}
 c_{20} &= - \int \der q\, p_5(q)\int\der\dot q\,  p(\dot{q}) \frac{f_{q\dot{q}}(s,q,\dot{q})}{f(s)}\notag \\
&\ph{-\int\der q}
\times\bigg(L_1^{(3/2)} (5 q^2/2)+5\Gamma^2 \sqrt{s}q\dot{q}\notag\\
&\ph{-\int\der q \times L_1^{3/2}(q^2)}
+ \Gamma^2L_1^{(-1/2)} (5 \Gamma^2 s \dot{q}^2/2)\bigg)\,,
\label{eq:c20}
\end{align}
where $\dot{q} \equiv \mathrm{d} q/ \mathrm{d} \sqrt{s}$, $f_{q\dot{q}}(s,q,\dot{q})$ is given by \eqn{eq:fsqqdot}, and $f(s)$ is  the first crossing distribution obtained by integrating $f_{q\dot{q}}(s,q,\dot{q})$ over $q$ and $\dot q$ (equation~\ref{eq:fs}).

Notice that, since $Q_5$ correlates with $\dot{q}$, large scale nonlocal bias also receives contributions from $\dot{q}$ and $\dot{q}^2$. In Figure \ref{fig:c2s}, the full analytic $c_2$ at two different scales (equation~\ref{eq:c2App}) is shown as a continuous line, and we see that it provides a good description of our Monte Carlo estimates of \eqn{eq:c2_hat}, shown as the green and yellow points. Although the measurements at the two scales are too close to each other to allow us an estimate of the scale dependent piece, they are still on sufficiently small smoothing scales that we can appreciate the difference between the points and $c_{20}$ (shown as the dashed line).

\begin{figure}
 \centering
 \includegraphics[width=0.9\hsize]{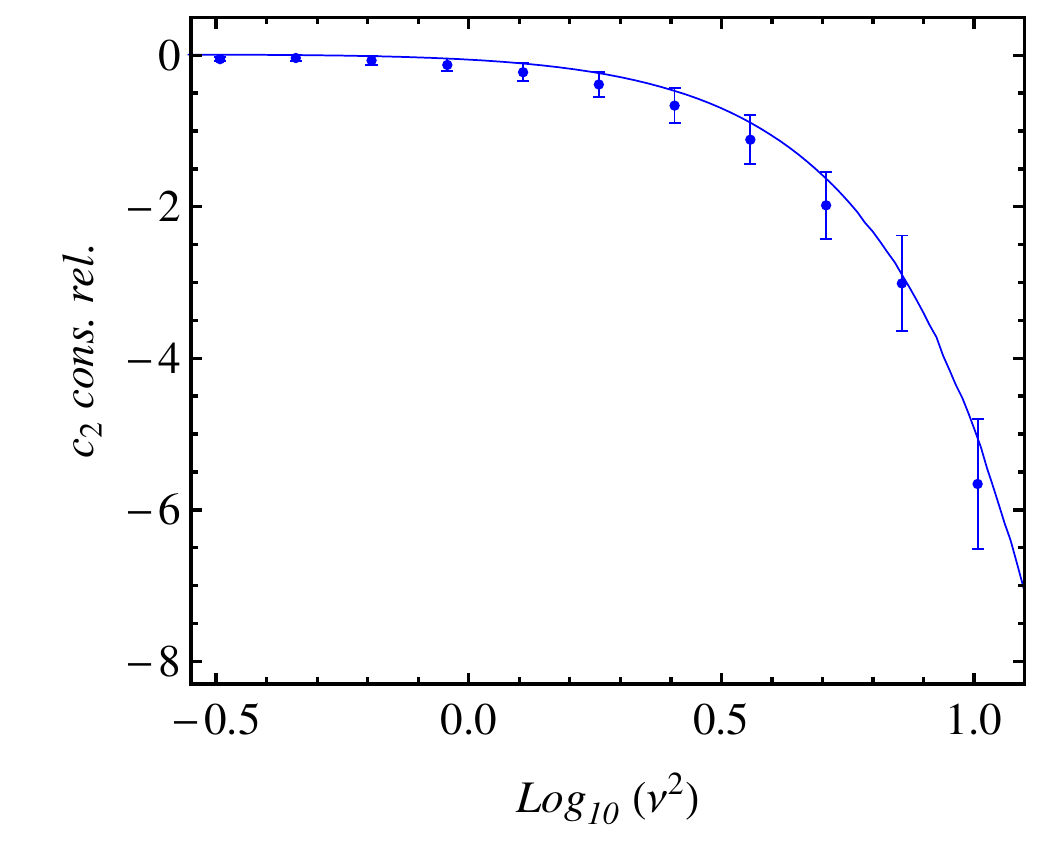}
 \caption{Comparison between the analytic prediction of nonlocal bias parameters $c_{2}$ (smooth curve) and the combination of moments of $\delta_{1\times}$ which appears on the rhs of equation~(\ref{eq:cons_c2}) (symbols). }
 \label{fig:cons_c2_ratio}
\end{figure}

We have already checked that moments of the PDF of the density at first crossing are well reproduced by the consistency relation for the $b_n$'s (Figures~\ref{fig:bias2dcq} and~\ref{fig:b2q->dcq}).  To close the loop, Figure~\ref{fig:cons_c2_ratio} tests the consistency relation for shear bias in the form of equation~(\ref{eq:cons_c2}). The agreement between the analytical prediction from the $c_2$ terms (smooth curve) and the particular combination of moments of $\delta_{1\times}$ given on the rhs of equation~(\ref{eq:cons_c2}) (symbols) is impressive; it supports our understanding of why density and shear bias coefficients are related, and how they can be used to study halo formation.  Presumably, good approximations to the higher order moments of $p(q_{1\times}|s)$ can also be written entirely in terms of the density bias factors $b_n$, but we have not pursued this further.

\subsection{Relation to previous work}
We now turn to the other two sets of symbols, the blue and red points, in Figure~\ref{fig:c2s}.  These show the result of replacing 
\be
 Q^2\to\zeta^2_5 \qquad {\rm and}\qquad 
 r \to r_{\zeta} = \frac{S_{1\times}}{\sqrt{s S_2}}
\label{eq:r}
\ee
in equation~(\ref{eq:c2_hat}), with $\zeta^2_5$ defined as follows.  Since $Q^2$ is $\chi^2$ with five degrees of freedom, we can think of it as $\sum_{i=1}^5 G_i^2/S$ where each of the $G_i$ are independent Gaussian random numbers.  We then set $Y^2\equiv\sum_{i=1}^5 Y_i^2$, where $Y_i\equiv \der G_i /\der S$, and finally define $\zeta^2_5\equiv Y^2/\langle Y^2\rangle$.  In this sense, $\zeta^2$ is to $Q^2$ as $\der\delta/\der s$ was to $\delta$ in the previous section.  For this reason, we might expect the result of averaging $L_1^{(3/2)}(5\zeta^2_5/2)$ around first crossing positions to be similar to, or (in the limit of infinite smoothing scale) the same as averaging $L_1^{(3/2)}(5Q^2_5/2)$, \emph{provided} we use the correct cross-correlation coefficient, $r_\zeta$, of \eqn{eq:r}.  The agreement between $\widehat{c}_{2,q_5}$ and $\widehat{c}_{2,\zeta_5}$, blue and red points, in Figure~\ref{fig:c2s} shows that this is indeed the case. As before, we are actually able to understand the small differences between the two set of measurements, since they arise by replacing $\epsilon_{\Delta\delta}$ in equation \ref{eq:c2App} with $\epsilon_{X\delta}$.

This agreement is particularly relevant to recent work on the role of shape in determining protohalo formation.  \cite{bcdp14} measured cross-correlations between protohaloes and a quantity which is like our $\zeta^2_5$.  Although they differ in detail -- their quantity is the same as ours only for a Gaussian smoothing filter -- the essential point is that the quantity they consider is correlated with $q^2_5$.  They define
\be
 s_2 \chi_{01} \equiv - \frac{1}{\tilde{r}^2} \frac{1}{N} \sum_{\alpha=1}^N L_1^{(3/2)}(5\zeta^2_{5,\alpha}/2) ,
\ee
which they expected would return the correlation between the small and large scale values of $\zeta$, for which the cross correlation coefficient is $\tilde{r}^2\equiv S_{2\times}^2/s_2S_2$.  Their prediction -- which is shown as a blue dashed curve in their Figure~5\footnote{Beware, the notation $\chi_{10}$ in \cite{bcdp14} is for the bias parameter obtained from cross-correlating with a quantity which does not correlate with $q_5$, so it has nothing to do with $c_{20}$!} 
 -- increases weakly as protohalo mass increases.  In contrast, their measurements -- shown as triangles in their Figure -- decrease strongly.  This discrepancy led them to question their assumption about the role halo shape plays in determining where haloes form.  

However, shape and shear are closely related; crudely speaking, shear is to shape as $\delta$ is to $\der\delta/\der s$.  Therefore, if shear plays a role in halo formation, then cross-correlating the large scale shape field with protohalo positions may actually return the bias factor associated with shear $c_{20}$ rather than shape.  
The agreement between $\widehat{c}_{2,q_5}$ and $\widehat{c}_{2,\zeta_5}$ in Figure~\ref{fig:c2s} shows that this is indeed what happens in our idealized toy model.  Since our $\zeta_5$ is very close to $\zeta_5$ in \cite{bcdp14}, their 
\be
 s_2\chi_{01} \approx (r_\zeta/\tilde{r})^2\,c_{20}
  = (S_{1\times}/S_{2\times})^2\,(s_2/s)\,c_{20}.
\ee  
The difference in correlation coefficient factors $r$ is a consequence of their assuming that shape is fundamental, whereas our model has shear as the fundamental quantity.  The prefactor on the right hand side decreases as $\nu$ increases, so it flattens the trend we see in Figure~\ref{fig:c2s}, making it more like that in Figure~5 of \cite{bcdp14}.  And, vice-versa, we have checked that this factor transforms their dashed curve into our solid one here.  This strongly suggests that the procedure used by \cite{bcdp14} actually returns the bias associated with shear rather than shape.  (This argument does not explain why their measured signal crosses zero at small masses.  In our walks, this happens if our sample size is not sufficiently large, but it is not clear if a similar effect is to blame for them as well.  For instance, it may be that the model we are using to illustrate the effect of shear on halo formation is too simplistic.  But building a more sophisticated model is beyond the scope of the present work.)

\section{Conclusions}
\label{sec:conclude}

We argued that it was useful to generalize the notion of bias with respect to the {\em overdensity} field of the dark matter, to the difference between averages around special positions in space and those over all space.  This makes it clear that averages centered on protohalo patches -- i.e. cross-correlations between biased tracers and the dark matter -- pick up correlations with any quantity which matters for halo formation.  Since bias is typically measured using averages on scales larger than those of a halo, the cross-correlation signal is typically dominated by the terms with the fewest orders of $k$.  As a result, cross-correlating protohaloes with the large scale density and curvature fields should return the same answer.  We showed this was true using the excursion set approach, in which, for many cross-correlations, the answer is known analytically (Figure~\ref{fig:b1s}).  We then used this fact to make two points.  

First, we showed that this allows one to estimate bias factors with no prior knowledge of the physics of collapse (Figure~\ref{fig:b10}).  In effect, this enables the cross-correlation methods pioneered by \cite{mps12} to be performed with fewer assumptions than before.  We illustrated this explicitly for linear and quadratic density bias factors (Figures~\ref{fig:b1s} and~\ref{fig:b2n}).  

We then showed that this same effect matters for other variables which are not correlated with overdensity.  One such pair of current interest is the quadrupolar pair associated with the tidal shear and shape.  
We argued that cross-correlating protohaloes with the large scale shape field will produce a signal which is very similar to that obtained by cross-correlating with the shear field (Figure~\ref{fig:c2s}).  This may explain some puzzling results in previous work \citep{bcdp14}; in particular, the puzzling cross-correlations there may actually be consistent with the nonlocal bias signals reported in \cite{css13} and \cite{scs12}.  

To date, it has been thought that one must know something about the physics of halo formation -- such as the criticial density $\delta_c(s)$ a protohalo patch must have if it is to become a halo of mass $m(s)$ -- to correctly predict halo abundances and bias.  We have shown that one can turn the argument around:  One can obtain interesting constraints on this physics from measurements of halo bias.  Our methodology correctly reproduces $\delta_c(s)$ with no prior knowledge of either the stochasticity associated with halo formation, its dependence on halo mass, or the mass and scale dependence of halo bias (Figures~\ref{fig:bias2dc} and~\ref{fig:bias2dcq}).  The same methodology can be used to quantify the stochasticity in halo formation (Figures~\ref{fig:b2todc} and~\ref{fig:b2q->dcq}), and to relate the nonlocal bias factors associated with this stochasticity to nonlinear density bias (equation~\ref{eq:cons_c2}).  This has the potential to vastly simplify analyses in the next generation of datasets which will be sensitive to nonlocal bias.  

In a companion paper \citep{cphs17}, we show how to extend some of the results discussed in this paper for the stochastic barrier which depends on the shear to the more realistic case where haloes are also peaks of the density field \citep{bbks,ps12b,psd13}. A comparison of the predictions of this new model to measurements of the halo mass function and the density and shear bias parameters in $N$-body simulations is also the subject of work in progress.

Finally, although our results relate to the Lagrangian bias of protohalo patches,  in the peaks model, the lowest order mapping to the Eulerian bias of the evolved halo field preserves the consistency relations we have exploited \citep{desjacques/crocce/etal:2010}. Work in progress studies if this is generic.

\section*{Acknowledgements}
EC, AP and RKS thank E. Sefusatti and L. Guzzo for their kind hospitality at INAF-OAB Merate during the summer of 2014. The research of AP is supported by the Associateship Scheme of ICTP, Trieste and the Ramanujan Fellowship awarded by the Department of Science and Technology, Government of India.

\bibliography{bibliography}

\appendix

\section{Cross-correlating with Laguerre polynomials in the stochastic model}
\label{App:A}
The main text described the motivation for computing cross correlations between special positions (protohaloes) and the large scale shear field:  since the shear field is $\chi^2_5$-distributed, the generalized Laguerre polynomials $L_n^{(3/2)}(x)$ are particularly interesting.  Here we show what this cross-correlation is expected to yield in the context of the stochastic model for halo formation described in the main text.  

Our procedure will be to show the result of averaging the generating function of generalized Laguerre polynomials,
\be
\label{eq:GenLag}
 \mathcal{L}(t|1+\alpha,x)\equiv \sum_m^\infty t^m\,L_m^{(\alpha)}(x)
 = \frac{{\rm e}^{-tx/(1-t)}}{(1-t)^{1+\alpha}},
\ee
over the special positions, from which the individual averages can be obtained in the usual way (by taking derivatives with respect to $t$).  This follows \cite{cs13} who studied the case when the stochastic variable is Gaussian rather than $\chi^2$.  
Before proceeding to the derivation we need to define a few quantities. From \eqn{eq:defq2} we define the derivative of $q$ with respect to $\sqrt{s}$
\begin{align}
\frac{\der q}{\der \sqrt{s}} \equiv \dot{q} &=  \frac{1}{5 q}\sum_i \frac{g_i}{\sqrt{s}} \left(\frac{\dot{g_i}}{\sqrt{s}}-\frac{g_i}{s}\right) \notag\\
&=  \frac{\Gamma \sqrt{s}}{5 q}\sum_i G_i X_i
\end{align}
which is Gaussian distributed with zero mean and variance $5/(\Gamma^2s)$ with $\Gamma^2\equiv\gamma^2/(1-\gamma^2)$ \citep{ms14}. For ease of notation we have defined the normalized variables $G_i = g_i/\sqrt{s}$ and $X_i = \dot{G_i} \Gamma\sqrt{s}$, which have the convenient property of being independent from each other, i.e. $\avg{G_iX_j}=0$ for all $i,j$. To obtain an expression for the bias coefficient with respect to traceless shear we have to compute the average Laguerre polynomial in the large scale shear $Q^2$ given some conditions on the small scale shear and its derivative, $\Cal{C}=\mathcal{C}(q,\dot{q})$,
\begin{align}
& \left\langle L_{1}^{(3/2)}\left(5 Q^2/2\right) \Big|\, \mathcal{C} \right \rangle = \notag\\
& \prod_i  \int p(G_i) \der G_i \; \prod_i \int p(X_i) \der X_i \; \prod_i \int p(G_{0,i} | G_i,\,X_i ) \der G_{0,i}   \notag \\
& \times \int \der Q^2 \,\dir\left(Q^2 - \sum_i G_{0,i}^2/5\right)  L_{1}^{(3/2)}( 5 Q^2/2) \,\Cal{C}(q,\dot{q}) 
\label{eq:bs2}
\end{align}
where $Q^2 = \sum G_{0,i}^2/5$ is the large scale shear. Throughout this section, we will assume that the constraint is suitably normalised, such that $\int\der q^2\,p(q^2)\int\der\dot q\,p(\dot q)\,\Cal{C}(q,\dot q)=1$. The more general excursion set constraint is discussed in Appendix~\ref{app:EScon}. 

The conditional probability of the large scale $G_{0,i}$ given the small scale $G_i$ and $X_i$ is a Gaussian with mean $r_G G_i + r_X X_i$ and variance $1-r_G^2-r_X^2$,  the cross-correlation coefficients being
\be
r_{G} \equiv \langle G_0 G \rangle = S_\times/\sqrt{sS}
\ee
and 
\be
r_{X} \equiv \langle G_0 X \rangle = \frac{S_\times\Gamma}{\sqrt{sS}}(\epsilon_{\Delta\delta}-1) = r_G \Gamma(\epsilon_{\Delta\delta}-1)\,,
\ee
where $\epsilon_{\Delta\delta}$ was defined in \eqn{epsilon}. The generating function in \eqn{eq:GenLag} allows us to rewrite \eqn{eq:bs2} as
\begin{align}
&\left \langle L_{1}^{(3/2)}\left(5 Q^2/2\right) \Big|\, \mathcal{C} \right  \rangle \notag \\
&=\prod_i  \int p(G_i) \der G_i \; \prod_i \int p(X_i) \der X_i \,\prod_i \int p(G_{0,i} | G_i,\,X_i ) \der G_{0,i}  \notag \\
&\ph{\prod_i  \int} 
\times \left[\cfrac{\partial}{\partial t} \mathcal{L}\left(t\Big| 5/2, \sum_i G_{0,i}^2/2\right) \bigg|_{t = 0}\right] \mathcal{C}(q,\dot{q}) \,,
\label{app:L13/2_genfunc}
\end{align}
which we then integrate over the large scale $G_{0,i}$ before taking the derivative. The net result is 
\begin{align}
& \left\langle L_{1}^{(3/2)}\left(5 Q^2/2\right)\Big|\, \mathcal{C} \right \rangle \notag\\
&= \prod_i  \int p(G_i) \der G_i \;\prod_i \int p(X_i) \der X_i   \notag \\
&\ph{\prod_i  \int}
\times\bigg[r_G^2L_{1}^{(3/2)}(\sum_i G_i^2/2) + r_X^2 L_1^{(3/2)}(\sum_i X_i^2/2) \notag\\
&\ph{\prod_i  \int[r_G^2L_{1}]}
-r_G r_X \sum_i G_i X_i \bigg]\,\mathcal{C}(q,\dot{q}) 
\label{eq:avgL13/2}
\end{align}
which explicitly shows that the integral is zero if $\mathcal{C}(q,\dot{q})  = 1$. The first and last terms in square brackets in \eqn{eq:avgL13/2} can be straightforwardly written in terms of $q$ and $\dot{q}$. The second term instead looks more complicated.

To proceed, it is useful to note the following identity for $\chi^2$ variables.  Namely, if $\eta^2\equiv \sum_{i=1}^n (\eta_i^2/s)/n$ and $\xi^2\equiv \sum_{i=1}^n (\xi_i^2/s)/n$ are independent $\chi^2_n$ variates, then the quantity $\sum_i (\eta_i\xi_i/s)/n$, being the dot product of the two underlying Gaussian vectors, can be written as $\eta\,\xi\cos\theta$.  The joint distribution of $g$, $\xi$ and $\theta$ is given by the product of three independent distributions:   
\begin{align}
 &p(\eta)\,p(\xi)\,p(\theta)\, {\rm d}\eta\,{\rm d}\xi\,{\rm d}\theta \nonumber\\
 &= 2\frac{{\rm d}\eta}{\eta}\,\frac{(n\eta^2/2)^{n/2}}{\Gamma(n/2)}\,{\rm e}^{-n\eta^2/2}
 2\frac{{\rm d}\xi}{\xi}\,\frac{(n\xi^2/2)^{n/2}}{\Gamma(n/2)}\,{\rm e}^{-n\xi^2/2}
 \nonumber\\
 &\ \times \ {\rm d}\theta\,\frac{\sin^{n-2}\theta}{B[1/2,(n-1)/2]}
\end{align}
where $B(a,b) = \Gamma(a)\Gamma(b)/\Gamma(a+b)$ is the Beta-function.  
It is useful to think of $\xi$ as a radial variable made from the Cartesian variables $\xi_x = \xi\cos\theta$ and $\xi_y=\xi\sin\theta$, so that ${\rm d}\xi_x\,{\rm d}\xi_y = {\rm d}\theta\,{\rm d}\xi\,\xi$.  In these variables, 
\begin{align}
 & {\rm d}\xi_x\,p(\xi_x)\, {\rm d}\xi_y\,p(\xi_y) 
   = p(\xi)\,p(\theta)\, {\rm d}\xi\,{\rm d}\theta \nonumber\\
 & = {\rm d}\xi_x\,\frac{{\rm e}^{-n\xi_x^2/2}}{\sqrt{2\pi/n}}\,
 2\frac{{\rm d}\xi_y}{\xi_y}\,\frac{(n\xi_y^2/2)^{(n-1)/2}}{\Gamma[(n-1)/2]}\,
  {\rm e}^{-n\xi_y^2/2}.
\end{align}
This shows that $\xi_x$ is a Gaussian with variance $1/n$, whereas $\xi_y$ is $\chi^2_{n-1}$.  

By matching notation, it is easy to see that $q$ here is $\eta$, and since $q\dot{q}$ is like a dot product, $\dot{q}$ is like $\xi_x/(\Gamma \sqrt{s})$.  \cite{ms14} noted that $p(\dot{q}|q)$ is a Gaussian, independent of $q$; our transformation from $(\xi,\theta)$ to $(\xi_x,\xi_y)$ shows why. 
Since the constraint $\mathcal{C}(q,\dot{q})$ depends on $q$ and $\dot{q}$ but not on $\xi_y$, and the distribution of $\xi_y$ is independent of $q$ and $\dot{q}$, we may integrate it out. 
We therefore arrive at the final expression for the average Laguerre given the constraint,
\begin{align}
\label{eq:Lag_q2}
&\left\langle L_{1}^{(3/2)}\left(5 Q^2/2\right) \Big|\,\mathcal{C}  \right \rangle   \notag \\
&= r_G^2\,\int \der q^2\, p(q^2)\;\int \der\dot{q}\,p(\dot{q}) \;  \mathcal{C} (q,\dot{q})\, \tilde{c}_2(q,\dot q;s,S)  \,,
\end{align}
where we defined
\begin{align}
\tilde{c}_2(q,\dot q;s,S) &\equiv L_{1}^{(3/2)}(5 q^2/2)-5  \Gamma^2\sqrt{s}(\epsilon_{\Delta\delta}-1) q\dot{q}  \notag \\ 
&\ph{L_1()}
+  \Gamma^2(\epsilon_{\Delta\delta}-1)^2  L_1^{(-1/2)}(5 \Gamma^2 s \dot{q}^2/2)   \,,
\label{eq:c2qqdot}
\end{align}
with the notation reminding us that this quantity depends on both the large and small scales.
It is convenient to write the expression for non local bias collecting terms by powers of the derivative of $q$, i.e. by powers of $\epsilon_{\Delta\delta}$. We thus write, similarly to $b_2$ in \eqn{eq:b2},
\begin{align}
\label{eq:c2App}
c_2 &= -r_G^{-2}\,\left\langle L_{1}^{(3/2)}\left(5 Q^2/2\right) \Big|\,\mathcal{C}  \right \rangle \notag\\
&=  c_{20} + 2\,\epsilon_{\Delta\delta}\,c_{21} + \epsilon_{\Delta\delta}^2\,c_{22}
\end{align}
where the coefficients $c_{2j}$ can be read off from \eqns{eq:Lag_q2} and~\eqref{eq:c2qqdot}:
\begin{align}
c_{22} & = - \Big\langle  \Gamma^2 L_1^{(-1/2)}(5 \Gamma^2 s \dot{q}^2/2) \Big| \, \mathcal{C} \Big\rangle 
\label{eq:c22App}\\
c_{21} &=  \Big\langle  5 \Gamma^2 \sqrt{s}\, q\dot{q}/2\,\Big| \, \mathcal{C} \Big\rangle -c_{22}
\label{eq:c21App}\\
c_{20}  &=   -\Big\langle L_{1}^{(3/2)}(5 q^2/2)\,\Big| \, \mathcal{C} \Big\rangle -2\,c_{21} - c_{22}\,.
\label{eq:c20App}
\end{align}
This shows that the consistency relation for shear, \eqn{eq:consrelq2} in the main text, holds. 

Although we have worked at lowest order in shear bias so far, the results above can be generalized to arbitrary order by working directly with the Laguerre generating function. For completeness, we sketch this calculation next. The basic idea is to perform the constrained average of the generating function before taking the derivative in \eqn{app:L13/2_genfunc}. This leads to
\begin{align}
&\left \langle \mathcal{L}\left(t|5/2,5Q^2/2\right) \Big|\, \mathcal{C} \right  \rangle \notag \\
&=\prod_i  \int p(G_i) \der G_i \; \prod_i \int p(X_i) \der X_i \,\prod_i \int p(G_{0,i} | G_i,\,X_i ) \der G_{0,i}\notag \\
&\times\int\der Q^2\dir\left(Q^2-\sum_i G_{0,i}^2/5\right) \mathcal{L}\left(t\Big| 5/2, 5Q^2/2\right) \mathcal{C}(q,\dot{q}) \notag\\
&=\prod_i  \int p(G_i) \der G_i \; \prod_i \int p(X_i) \der X_i \,\prod_i \int p(G_{0,i} | G_i,\,X_i ) \der G_{0,i}\notag \\
&\ph{\prod_i  \int} 
\times \mathcal{L}\left(t\Big| 5/2, \sum_i G_{0,i}^2/2\right) \mathcal{C}(q,\dot{q}) \,.
\end{align}
The integral over the $G_{0,i}$ is a straightforward convolution of Gaussians, leading to
\begin{align}
&\prod_i \int p(G_{0,i} | G_i,\,X_i ) \der G_{0,i} \,\mathcal{L}\left(t\Big| 5/2, \sum_i G_{0,i}^2/2\right)\notag\\
&=\prod_i\frac{\exp\left(-t\left[r_G\,G_i + r_XX_i\right]^2/2t\Sigma^2\right)}{(t\Sigma^2)^{1/2}}
\label{app:<Laggen>}
\end{align}
where
\be
t\Sigma^2 = 1-t\left(r_G^2+r_X^2\right)\,.
\ee
The evaluation of $\left \langle \mathcal{L}\left(t|5/2,5Q^2/2\right) \Big|\, \mathcal{C} \right  \rangle$ is therefore similar to that of $\left \langle L_1^{(3/2)}\left(5Q^2/2\right) \Big|\, \mathcal{C} \right  \rangle$ in \eqn{eq:avgL13/2}, with the expression in square brackets in that equation replaced by the r.h.s. of \eqn{app:<Laggen>}. The discussion below \eqn{eq:avgL13/2} carries through, with the result
\begin{align}
&\left \langle \mathcal{L}\left(t|5/2,5Q^2/2\right) \Big|\, \mathcal{C} \right  \rangle \notag \\
&= \int \der q^2\, p(q^2)\int \der\dot{q}\,p(\dot{q}) \;  \frac{{\rm e}^{-t5\Cal{Q}^2/2(t\Sigma^2)}\,\mathcal{C} (q,\dot{q}) }{\left(t\Sigma^2\right)^{1/2}\left(1-tr_G^2\right)^{3/2}}\,,
\label{app:<Laggen>-final}
\end{align}
where we defined
\be
\Cal{Q}^2 \equiv \left(r_Gq + r_X\Gamma\sqrt{s}\,\dot q\right)^2/\left(r_G^2+r_X^2\right)\,.
\ee
Equation~\eqref{app:<Laggen>-final} can now be differentiated $j$ times with respect to $t$ to obtain bias parameters $c_{2j}$ of arbitrary order.

\section{Details of calculations in stochastic models}
\label{app:EScon}
In this Appendix, we provide some formulae for the first crossing distribution $f(s)$. 
Following \cite{ms14}, if $q^2$ is drawn from a $\chi_5$ distribution obtained by summing $5$ independent Gaussian variables with variance $s$, the up-crossing distribution of the process in \eqn{dcq} can be written as 
\be
\label{eq:fs}
f(s) = \int \der q\, p_5(q) \int\der\dot q\, p(\dot{q})\, f_{q\dot{q}}(s,q,\dot{q};q_c)
\ee
where $p_5(q)$ is the $\chi$-distribution associated with $q^2$ and $p(\dot{q})$ is a Gaussian with zero mean and variance $\langle\dot{q}^2\rangle = 5 / \Gamma^2 s$. The function $f_{q\dot{q}}(s,q,\dot{q};q_c)$ is given by
\begin{align}
&s f_{q\dot{q}}(s,q,\dot{q};q_c) \notag\\
&\ph{s}
=\frac{e^{- B^2/2s}}{\sqrt{2 \pi}}  \int\frac{\der x\,{\rm e}^{-\left(x-\gamma B/\sqrt{s}\right)^2/2(1-\gamma^2)}}{\sqrt{2\pi(1-\gamma^2)}}\notag\\
&\ph{sfs\int\der x\,p_{\rm G}(x)}
\times \left(x/\gamma-\dot B\right)\,\HT\left(x/\gamma-\dot B\right) \,,
  \label{eq:fsqqdot}
\end{align}
where $B = \delta_c + (\delta_c/q_c)\,\sqrt{s}\,q$ and $\dot B = \der B/\der\sqrt{s} = (\delta_c/q_c)(q+\sqrt{s}\dot q)$. The integral over $\dot q$ in \eqn{eq:fs} can then be performed analytically, leaving two integrals (over $x$ and $q$) that must be performed numerically. For brevity, we omit the expression involving these two integrals.

The smooth curve in Figure~\ref{fig:fs} shows that this formula provides an excellent description of the first crossing distribution in Monte Carlo realizations of the process. 
The expressions for non local bias we derived in the previous section follow through for the more general excursion set constraint discussed here, simply by replacing the constraint $\mathcal{C} (q,\dot{q})$ with the excursion set one, \eqn{dcq}, since the latter does not depend on $\xi_y$. 
This is equivalent to replacing $\mathcal{C} (q,\dot{q}) \to f_{q\dot{q}}(s,q,\dot{q};q_c)/f(s)$ in \eqn{eq:Lag_q2}, which gives
\begin{align}
\label{eq:Lag_q2ES}
&\Big\langle L_{1}^{(3/2)}\left(5 Q^2/2\right) \Big| \text{Exc. Set}  \Big\rangle   \notag \\
&= r_G^2\,\int \der q\, p_5(q)\int\der\dot q\, p(\dot{q})\, \frac{f_{q\dot{q}}(s,q,\dot{q};q_c)}{f(s)}
\, \tilde{c}_2(q,\dot q;s,S)\,,
\end{align}
where $\tilde{c}_2(q,\dot q;s,S)$ was defined in \eqn{eq:c2qqdot}.

\label{lastpage}

\end{document}